%% file: main.tex
\documentclass[lettersize,journal]{IEEEtran}
\usepackage{amsmath,amsfonts}
\usepackage{array}
\usepackage[caption=false,font=normalsize,labelfont=sf,textfont=sf]{subfig}
\usepackage{textcomp}
\usepackage{stfloats}
\usepackage{url}
\usepackage{verbatim}
\usepackage{graphicx}
\usepackage{cite}

\usepackage{graphicx} 
\usepackage{amsmath,amssymb}
\usepackage{multicol}
\usepackage{booktabs}
\usepackage{algorithm}
\usepackage{algpseudocode}
\usepackage[dvipsnames]{xcolor}

\usepackage{fancyhdr}
\fancypagestyle{firstpage}{%
\chead{\small This article has been accepted for publication in IEEE Transactions on Wireless Communications. This is the author's version which has not been fully edited and
content may change prior to final publication. Citation information: DOI 10.1109/TWC.2024.3450190}

\cfoot{\small \textcopyright~2024 IEEE. Personal use is permitted, but republication/redistribution requires IEEE permission.
See https://www.ieee.org/publications/rights/index.html for more information.}
}

\begin{document}

\title{Uplink MIMO Detection using Ising Machines: A Multi-Stage Ising Approach}

\author{\IEEEauthorblockN{Abhishek~Kumar~Singh$^{1}$, Ari Kapelyan$^{4}$, Minsung Kim$^{1,2}$, Davide~Venturelli$^{3}$, Peter L. McMahon$^4$, and Kyle~Jamieson$^1$}\\
\IEEEauthorblockA{$^1$\textit{Department of Computer Science, Princeton University}}\\
\IEEEauthorblockA{$^2$\textit{Department of Computer Science, Yale University}\\$^3$\textit{USRA Research Institute for Advanced Computer Science}}\\$^4$\textit{School of Applied and Engineering Physics, Cornell University}}

\maketitle
\thispagestyle{firstpage}

\begin{abstract}
Multiple-Input-Multiple-Output~(MIMO) signal detection is central to nearly every state-of-the-art communication system, and enhancements in error performance and computational complexity of MIMO detection would significantly enhance data rate and latency experienced by the users. Theoretically, the optimal MIMO detector is the maximum-likelihood (ML) MIMO detector; however, due to its extremely high complexity, it is not feasible for large real-world communication systems. Over the past few years, algorithms based on physics-inspired Ising solvers, like Coherent Ising machines and Quantum Annealers, have shown significant performance improvements for the MIMO detection problem. However, the current state-of-the-art is limited to low-order modulations or systems with few users. In this paper, we propose an adaptive multi-stage Ising machine-based MIMO detector that extends the performance gains of physics-inspired computation to Large and Massive MIMO systems with a large number of users and very high modulation schemes~(up to 256-QAM). We enhance our previously proposed delta Ising formulation and develop a heuristic that adaptively optimizes the performance and complexity of our proposed method. We perform extensive micro-benchmarking to optimize several free parameters of the system and evaluate our methods' BER and spectral efficiency for Large and Massive MIMO systems (up to 32 users and 256-QAM modulation). 
\end{abstract}

\IEEEpeerreviewmaketitle

\let\thefootnote\relax\footnote{The work of Abhishek Kumar Singh was supported by the National Science Foundation~(NSF) under Award CNS-1824357. The work of Ari Kapelyan was supported by NSF under Award CCF-1918549. The work of Minsung Kim was supported by NSF under Awards CNS-2147946, CNS-2216030, and CNS-2211945. The work of Davide Venturelli was supported by NSF under Award CCF-1918549. The work of Peter L. McMahon was supported by NSF under Award CCF-1918549. The work of Kyle Jamieson was supported by NSF under Award CNS-1824357.}
\input{introduction.tex}
\input{systemModel.tex}
\input{convergence}
\input{design.tex}
\input{eval.tex}

\input{conclusion.tex}

\bibliographystyle{IEEEtran}
\bibliography{IEEEabrv,ref}
\end{document}

%% file: introduction.tex
\section{Introduction}
Multiple-Input Multiple-Output~(MIMO) signal detection is a key task in nearly every modern wireless system. More than two decades of research have been dedicated to finding an efficient and practically feasible MIMO detector that can achieve good error performance. The optimal Maximum Likelihood~(ML) detector is known to be NP-Hard~\cite{mimoReview} and is therefore infeasible for practical systems with a large number of users. Over the years, researchers have proposed several MIMO detection algorithms and a tradeoff between complexity and optimality can be observed in the performance of these methods~\cite{mimoReview}.  

The invention of Massive MIMO~\cite{lundMM,megaMIMO} allowed practical systems to circumvent this complexity-optimality trade-off by having a larger number of base station antennas than user antennas. Such systems have extremely well-conditioned channels and thus even linear detectors like the Minimum Mean-Squared Error~(MMSE) detector could achieve near-optimal performance~\cite{mmseMassiveMimo1,mmseMassiveMimo3}. However, this complexity reduction is not without trade-offs: having fewer users than base station antennas reduces the maximum throughput achievable by the cell and the number of users that can be served simultaneously. Therefore, developing an efficient MIMO detection algorithm that can achieve near-optimal performance allows us to expand massive MIMO systems into large MIMO systems (which have same number of transmit and receive antennas) and potentially improve cell capacity~\cite {ri-mimo}.

Several conventional computation-based methodologies in the literature aim to achieve near-optimal MIMO detection with polynomial complexity. These methods rely on mathematical and algorithmic techniques including channel inversion, approximate tree-search~\cite{fcsd}, lattice reduction~\cite{latticeReduction}, successive interference cancellation~\cite{optimalMMSEsic}. However, their gains reduce drastically as the size of the MIMO system increases, or they require a significant increase in complexity to maintain performance, affecting their scalability for practical systems. This prompts us to explore alternative approaches for ML-MIMO detection that can provide scalable near-optimal MIMO detection for real-world systems. 

Over the last few years, the use of machine learning, variational inference, and physics-inspired computation for MIMO detection has attracted the attention of researchers. Variational Bayes~(VB) inference relies on iterative estimation of the posterior probability distribution of the transmitted signal. VB-based methods have shown promising performance gains for Massive/Large MIMO systems with lower modulation schemes (smaller than 64-QAM)~\cite{VB1,VB2,VB3,VB4,VB5} or MIMO systems with only a few users ($\leq 4$)~\cite{VB6,VB7,VB8,VB9}. 
Researchers have also explored Approximate Message Passing (AMP) and AMP-assisted machine learning methods for MIMO detection. AMP is an iterative algorithm for signal detection~\cite{oamp,oampTutorial} in linear systems. AMP and other message passing based methods have shown promising performance results for Massive MIMO systems, Large MIMO systems with a few users ($\leq 4$)  or modulations smaller than 64-QAM~\cite{AMP1,AMP2,AMP3,AMP4,MP1,MP2,MP3}. 
VB/AMP needs to achieve iterative convergence which adversely affects the feasibility of an efficient parallel implementation.  As a result, it remains unclear whether the gains demonstrated by VB/AMP-based methods will extend to real-world large MIMO systems that use high-order modulations and serve many users.  

 Several machine learning approaches have been proposed for signal detection in a MIMO system. Machine learning-based methods adopt several different approaches like clustering with Gaussian Mixture Models~\cite{ML5}, Convolutional Neural Networks~(CNN)~\cite{ML0}, Recurrent Neural Networks (RNN)/LSTMs~\cite{ML1,lisa}, Deep Neural Networks~\cite{detnet}, or augment popular legacy detection methodologies like Successive Interference Cancellation (SIC) using machine learning~\cite{MLsic}. These methods have demonstrated good performance gains for Massive MIMO scenarios or Large MIMO scenarios with few users/low order modulations. Large MIMO systems with high-order modulations are addressed by the OAMPNet-2~\cite{mlModeldriven} detector, which relies on model-driven deep learning for joint MIMO detection and channel estimation, and demonstrates promising gains with ideal Rayleigh fading channels (even for $32 \times 32$ Large MIMO system with 64-QAM modulation). However, it experiences significant performance deterioration on realistic channels and its performance becomes similar to~MMSE. 
 
 While exploring machine learning and variational inference-based solutions for MIMO detection in Large MIMO systems (with high order modulation and a large number of users) remains an interesting problem, in this paper, we focus on a radical alternative to conventional MIMO detectors: Physics-inspired computation. Physics-inspired computation serves as an alternative to conventional digital computation and has shown promising results for solving NP-Hard problems~\cite{isingMaxCut,isingGraphColor}. These physics-inspired methods rely on creating a physical system whose dynamics embed the problem coefficients of an NP-Hard problem, and its steady state can be used to find the solution to the original problem. Physics-inspired methodologies include Quantum Annealing~(QA)~\cite{hauke2020perspectives}, Coherent Ising machines~\cite{wang2013coherent,marandi2014network,mcmahon2016fully,inagaki2016coherent,dopo,oeo}, digital-circuit Ising solvers~\cite{yamaoka201520k,aramon2019physics,goto2019combinatorial,goto2021high,leleu2020chaotic}, bifurcation machines~\cite{tatsumura2021scaling}, photonic Ising machines besides CIMs~\cite{roques2020heuristic,prabhu2020accelerating,babaeian2019single,pierangeli2019large}, spintronic and memristor Ising machines~\cite{sutton2017intrinsic,grollier2020neuromorphic,cai2020power}, and oscillation-based Ising machines~\cite{oim}. Researchers have explored applications of physics-inspired computation for MIMO detection~\cite{review2021toappear,QAdeluna}, and have shown promising results. Several Quantum Annealing-based approaches, purely quantum\cite{minsung2019} or classical-quantum hybrid~\cite{pSuccessQA,kim2022warm}, have been proposed and have shown significant performance gains for MIMO detection when low-order modulations~(4-QAM and lower) are used. Researchers have also demonstrated performance improvements and near-optimal behavior for MIMO detection with other physics-inspired methodologies like Coherent Ising Machines~\cite{ri-mimo}, Oscillator-based Ising Machines~\cite{oimMuMIMO}, Quantum Approximate Optimization Algorithm~(QAOA)~\cite{qaoaMimo} and Parallel Tempering~\cite{minsungParallelTemp}. However, as seen from results in the existing literature, the performance gains of physics-inspired methods and Quantum Annealing-based methods are diminished when higher-order modulations~(16-QAM or higher) are used~\cite{ri-mimo,pSuccessQA,kim2022warm,minsungParallelTemp,minsungMag} as the probability of finding the ground state for 16-QAM modulation can be smaller than $10^{-3}$ in the case of QA~\cite{pSuccessQA}. Further, it has been shown (for the Max-cut problem) that, with an increase in problem size, the probability of finding the ground state reduces rapidly for QA compared to CIM, and can be several orders of magnitude lower for problems with more than 40 spin variables~\cite{cim_v_qa}. Existing Ising machine-based MIMO detectors, like the QuAMax method~\cite{minsung2019}, 
 Paramax~\cite{minsungParallelTemp} and Regularized Ising MIMO (RI-MIMO)~\cite{ri-mimo}, rely on a direct transformation (trivial) of the ML-MIMO problem into an Ising problem, followed by using a physics-inspired solver to obtain solutions. The direct transform expresses the entire discrete integer-valued search space via its binary representation (including the regions that are highly unlikely to contain the correct solution). The existing Ising machine-based MIMO detectors provide performance improvements only for lower modulations and low signal-to-noise-ratio~(SNR) scenarios, and there are two major limitations:
\begin{enumerate}
    \item Bit error rate (BER) floor at high SNR, \textit{i.e.}, BER saturates and doesn't reduce with increasing SNR.
    \item No performance gains for higher order modulations (16-QAM or higher). Note that, this is critical as practical wireless systems like LTE/5G widely use modulations up to 256-QAM.
\end{enumerate}
These limitations restrict the practical feasibility of these methods. The first problem was addressed via regularized Ising formulation~\cite{ri-mimo}: RI-MIMO utilizes a low-complexity approximate solution to add a scenario-(SNR, MIMO size, modulation) dependent regularisation term into the Ising problem. This method mitigates the error-floor issue and provides good performance gains across all SNRs. However, as noted before, RI-MIMO also relies on a direct transformation (trivial) to represent the entire discrete-valued search space and convert a MIMO problem into an Ising problem. While RI-MIMO addresses one of the problems, it is also not able to provide performance improvements for 16-QAM or higher modulations. In contrast, our proposed algorithm involves a specialized transformation that allows for adaptive selection of the size and the location of the search space while transforming the MIMO problem into an Ising form. Our proposed multi-stage approach optimizes the size and location of the search space to improve performance and, as we will show later, can provide significant performance improvements even for high-order modulations (16-, 64- and 256-QAM). 

The current state-of-the-art MIMO detection with non-conventional methodologies (MIMO detection with machine learning, variational inference, and physics-inspired computation) falls short in addressing Large MIMO systems with high-order modulations (64-QAM and higher) and many users~($\geq 10$). To the best of our knowledge, our proposed methodology is the only known physics-inspired MIMO detector that overcomes these limitations and can provide significant performance gains for Large MIMO systems with many users~($\geq 10$) and higher-order modulations (up to 256-QAM).

In this paper, we propose the Multi-Stage Delta Ising MIMO algorithm~(MDI-MIMO); MDI-MIMO concurrently searches around one or many approximate solutions while adaptively adjusting the search radius and re-centering the search space as it progressively improves the solutions.  

Our main contributions can be summarized as follows:
\begin{enumerate}
    \item We propose the degenerate Delta Ising (dDI) formulation which allows adaptive selection of the search space~(size and location) of the ML-MIMO problem while transforming it into an Ising optimization problem. Note that this is radically different from the direct transform used by the existing methodologies which rely on a trivial transformation of the entire discrete valued search space.
    \item We propose a heuristic-based approach to determine the suitable reduced search space for the ML-MIMO problem.
    \item We propose a novel multi-stage MIMO decoding approach that iteratively adjusts the search space and the initial guess for the Ising-machine-based optimization, leading to significant performance improvements (the existing methodologies simply search over the entire search space without imposing any constraints).
    \item We demonstrate that the gains of our proposed methods are not limited to Coherent Ising Machines but can also extend to other Ising optimization algorithms/solvers like Parallel Tempering.
    \item Unlike prior designs, our proposed design enables physics-inspired computation methods to perform MIMO decoding well with high-order modulations such as 64- and 256-QAM (the existing methodologies fail to provide performance gains for 16-QAM and higher modulations).   
\end{enumerate}

The rest of the paper is organized as follows: Section~\ref{sec:sysModel} describes the MIMO system model and the maximum likelihood MIMO detector. Section~\ref{sec:cim} is a primer on Coherent Ising machines~(CIM) and describes the model used to evaluate our methods.
Section~\ref{sec:design} describes our proposed degenerate delta Ising formulation, a heuristic approach for estimating the required search radius (based on the channel measurements and received signal), and our Multi-Stage Delta Ising MIMO algorithm. Section~\ref{sec:eval} presents our evaluation results. We evaluate the BER and spectral efficiency of our methods for ($16$ users $\times$ $16$ antennas), ($32$ users $\times$ $32$ antennas), ($16$ users $\times$ $32$ antennas), and ($32$ users $\times$ $64$ antennas) MIMO systems. Our empirical analysis shows that our methods can provide significant performance improvements over MMSE-SIC~\cite{optimalMMSEsic} and the commonly used MMSE detector: achieving a BER of $10^{-3}$ (with 16-QAM and 64-QAM modulations) at an SNR approximately 4~dB lower than MMSE-SIC and 18~dB lower than MMSE, and providing throughput improvements (for large MIMO systems) of up to 100\% over MMSE-SIC and up to 200\% over MMSE. Even in the case of massive MIMO systems, where MMSE and MMSE-SIC perform extremely well~\cite{mmseMassiveMimo1,mmseMassiveMimo3}, our methods provide a 50\%  throughput improvement. We compare our methods against several machine learning/AMP-based MIMO detectors and show that, for a ($4$ users $\times$ $4$ antennas) MIMO system, our proposed methods significantly outperform them and achieve near-optimal performance. We perform extensive micro-benchmarking experiments to optimize the free parameters of our algorithm and justify our design choices. We evaluate our methods for a realistic ($16$ users $\times$ $16$ antennas) LTE scenario and demonstrate that our performance gains extend to realistic MIMO systems. In Section~\ref{sec:parallelTempering}, we show that the performance gains of our proposed methods are not limited to Coherent Ising machines but also generalize to other physics-inspired methodologies like Parallel Tempering (PT)~\cite{minsungParallelTemp}. It is observed that with the same PT solver, over two orders of magnitude better BER is achieved for 16-QAM ($16$ users $\times$ $16$ antennas)~MIMO with our proposed design, compared to the baseline design. We conclude our work and discuss future directions in Section~\ref{sec:conclusion}.

%% file: systemModel.tex
\section{System Model}
\label{sec:sysModel}
Consider an uplink MIMO Scenario with $N_r$ antennas at the base station and $N_t$ users with one transmit antenna each. This is equivalent to a $N_t \times N_r$ MIMO system and is described by a complex-valued $N_r \times N_t$ channel matrix $\mathbf{\tilde{H}}$. The symbol transmitted by each user is drawn from a fixed set $\Phi$ representing the M-QAM constellation. Let the transmit vector be $\mathbf{\tilde{x}}$ ($\mathbf{\tilde{x}} \in \Phi^{N_t}$), then the received vector $\mathbf{\tilde{y}}$ is given by
\begin{equation}
    \mathbf{
    \tilde{y}} = \mathbf{\tilde{H}}\mathbf{\tilde{x}} + \mathbf{n},
\end{equation}
where $\mathbf{n}$ is the channel noise. If the channel noise is assumed to be white Gaussian noise, then the maximum likelihood estimate for the transmitted vector $\mathbf{\tilde{x}}$ is given by
\begin{equation}
    \mathbf{\tilde{x}_{ML}} = \arg \min_{\mathbf{\tilde{u}} \in \Phi^{N_t}} ||\mathbf{\tilde{y}}-\mathbf{\tilde{H}}\mathbf{\tilde{u}}||^2,
    \label{eq:ML-MIMO}
\end{equation}
where the optimization variable $\mathbf{\tilde{u}}$ represents a possible transmit vector and the optimization requires minimization over all possible transmit vectors. Note that $\mathbf{\tilde{x}}$, $\mathbf{\tilde{y}}$ and $\mathbf{\tilde{H}}$ are all complex valued. Using the transform described in~\cite{ri-mimo},
\begin{equation}
\mathbf{H} = 
  \left[ {\begin{array}{cc}
   \Re(\mathbf{\tilde{H}}) & -\Im(\mathbf{\tilde{H}}) \\
   \Im(\mathbf{\tilde{H}}) & \Re(\mathbf{\tilde{H}}) \\
  \end{array} } \right],\text{~}
\end{equation}
\begin{equation}
\mathbf{y} =
  \left[ {\begin{array}{c}
   \Re(\mathbf{\tilde{y}}) \\
   \Im(\mathbf{\tilde{y}}) \\
  \end{array} } \right],\text{~}\mathbf{x}=
  \left[ {\begin{array}{c}
   \Re(\mathbf{\tilde{x}}) \\
   \Im(\mathbf{\tilde{x}}) \\
  \end{array} } \right],
  \label{eq:realTransVec}
\end{equation}
we get an equivalent real-valued problem,
\begin{equation}
    \mathbf{x_{ML}} = \arg \min_{\mathbf{u} \in [\Re(\Phi)^{N_t},\Im(\Phi)^{N_t}]} ||\mathbf{y}-\mathbf{H}\mathbf{u}||^2.
    \label{eq:realML-MIMO}
\end{equation}
In this paper, our objective is to efficiently solve~(\ref{eq:realML-MIMO}) using a Coherent Ising Machine and get the solution to~(\ref{eq:ML-MIMO}) by inverting the transform~(\ref{eq:realTransVec}).
\section{Primer: Coherent Ising Machines}
\label{sec:cim}
An Ising optimization problem~\cite{isingNpHard1} is a quadratic, unconstrained, NP-Hard optimization problem:
\begin{equation}
    \arg \min_{\forall i,s_i\in \{-1,1\}}-\sum_{i}h_is_i -\sum_{i \neq j}J_{ij}s_{i}s_{j},
    \label{eq:IsingBase}
\end{equation}
where $h{_i}$ and $J_{ij}$ are real valued problem coefficients and each \textit{spin} variable $s_{i} \in \{-1,1\}$. While the Ising problem contains both linear and quadratic terms, the linear terms can be converted into quadratic terms by using a single auxiliary spin variable~\cite{ri-mimo}. The key idea is to multiply each linear term with the auxiliary spin variable $s_a$ ($h_is_i \rightarrow h_is_is_a$). It has been shown that the optimal solution of the original problem~(\ref{eq:IsingBase}) can be obtained from the optimal solution of this modified problem~\cite{ri-mimo}. Therefore, without any loss of generality, an Ising problem with $N_s$ spin variables can be expressed as:
\begin{equation}
    \arg \min_{\forall i,s_i\in \{-1,1\}}-\sum_{i \neq j}J_{ij}s_{i}s_{j}=\arg \min_{\mathbf{s}\in \{-1,1\}^{N_s}}  - \mathbf{s}^T\mathbf{J}\mathbf{s},
    \label{eq:Ising}
\end{equation}
where $\mathbf{s} = \{s_1,s_2,...s_{N_s}\} \in \{-1,1\}^{N_s}$ 
and the diagonal entries of the matrix $\mathbf{J}$ are zeros. Note that, discussing a connection between the ML problem and the physical interpretation of the Ising problem is beyond the scope of this work and several theoretical works in the literature explore this connection~\cite{theory1,theory2}.

In the abstract, we can think of an Ising machine as a system or algorithm that takes the coefficients of the Ising problem as the input and outputs a candidate solution to~(\ref{eq:Ising}). While the Ising machine aims to find the global optimum, it can return a sub-optimal solution. 

In this paper, we focus on Coherent Ising Machines~(CIMs), but our methods can also be applied to other Ising machines. Coherent Ising machines were originally proposed~\cite{marandi2014network,mcmahon2016fully,inagaki2016coherent} using an artificial optical spin network to find the ground state of the Ising problem. Coherent Ising machines rely on creating interactions between physical entities, for example, optical pulses~\cite{dopo}, and the strength of these interactions is related to the coefficients of the Ising problem ($J_{ij}$). It has been shown that the steady state of CIMs reflects the optimal solution of the corresponding Ising   problem~\cite{dopo}. While large-scale real implementations of CIMs are not currently present, it is possible to emulate the computation principles of these devices via the ordinary differential equations (ODE) describing their behavior. This can be used to create a computational model that implements these ODEs on CPU/FPGA and tries to extract the benefits of CIMs~\cite{ampCorrectionCIM,ri-mimo}.

In accordance with prior research~\cite{ampCorrectionCIM}, the dynamics of several bi-stable CIM systems can be modeled with real-valued variables $x_i$ such that the corresponding \textit{spin} $s_i = sign(x_i)$. Given the Ising optimization problem~(\ref{eq:Ising}), 
the time evolution of such a system can be modeled using the ODE~\cite{ampCorrectionCIM}
\begin{equation}
    \forall i\text{,      }\dfrac{dx_i}{dt} = (p-1)x_i - x_i^3 + \epsilon \sum_{j\neq i} J_{ij}x_j.
    \label{eq:cim-base}
\end{equation}

Note that the CIM model used by RI-MIMO~\cite{ri-mimo} can be approximated by~(\ref{eq:cim-base}) as well. However, the model described by~(\ref{eq:cim-base}) is susceptible to locally stable solutions~(local minima of the Lyapunov function) and limit cycles~\cite{ampCorrectionCIM}, which can severely affect the likelihood of finding the ground state of~(\ref{eq:Ising}). An enhanced model with an amplitude heterogeneity correction~(AHC)~\cite{ampCorrectionCIM} can destabilize the local minima and avoid limit cycles. The AHC-based model introduces new ``error" variables $e_i$:


\begin{flalign}
 \forall i \text{      ,}\dfrac{dx_i}{dt} & = (p-1)x_i - x_i^3 + \epsilon e_i \sum_{j\neq i} J_{ij}x_j  \label{eq:cim1}\\
  \forall i \text{,      } \dfrac{de_i}{dt} & = -\beta(x_i^2 - a)e_i,\text{    }e_i > 0 \label{eq:cim2},
\end{flalign}
where $a,p,\epsilon$, and $\beta$ are the free parameters of the model. We set $p = 0.98$~\cite{ampCorrectionCIM} and slowly ramping up $\epsilon$ with time $t$ ($\epsilon = \gamma t$). $a,\beta$, and $\gamma$ can be appropriately selected to maximize performance. We empirically choose $a = 2$, $\beta = 1$ and linearly ramp up $\epsilon$ from 0 to 1000. 

%% file: convergence.tex
\subsection{Convergence and Complexity of CIM Model}
\label{sec:ising_conv}
\begin{figure}
    \centering
    \includegraphics[width = \linewidth]{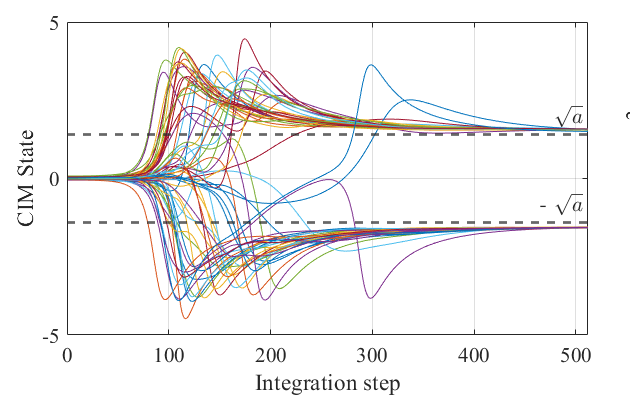}
    \caption{Example: Typical evolution of the CIM state denoted by variables $x_i$.}
    \label{fig:cim-trajectory1}
\end{figure}

Note that, at steady state, (\ref{eq:cim2}) implies that the state variables $x_i$ converge to approximately $\sqrt{a}$ or $-\sqrt{a}$. Integrating the CIM equations with the Range-Kutta method (RK4)~\cite{rk4}  with $dt = 0.005$, we observe that usually the CIM state, denoted by variables $x_i$, oscillates wildly before converging to a steady state value (as seen in Fig.~\ref{fig:cim-trajectory1}). As noted before, the sign of the state variables at steady-state is the spin output of the Ising machines.

\begin{figure}
    \centering
    \includegraphics[width = \linewidth]{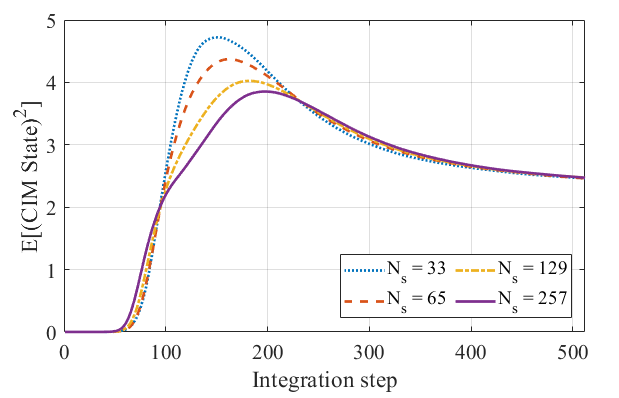}
    \caption{Convergence of CIM: variations of the expected value of the square of the CIM state variables demonstrate that approximately 500 integration steps are sufficient for the ODEs to achieve the steady-state behavior. The expectation is taken over different state variables and 1000 problem instances.}
    \label{fig:trajectory-conv}
\end{figure}

Given that the CIM is a bi-stable system, \textit{i.e.} at steady-state a CIM converges to two possible states, we look at the $E[x_i^2]$ (where the expectation is across different Ising instances and state variables) to determine the required number of iterations for numerical integration. We see from Fig.~\ref{fig:trajectory-conv}, where $N_s$ represents the number of spin variables in the Ising problem, that 512 integration steps are sufficient even for very large problems with more than 200 spin variables. While the length of the numerical integration can be potentially optimized depending on problem size and convergence behavior, in this paper, we fix the length of numerical integration to 512 steps and focus on the algorithmic aspects of the problem.  

For an Ising problem with $N_s$ spin variables, each integration step requires O($N_s^2$) operations. Given that we execute a fixed number of integration steps, the time complexity of integrating the CIM equations is given by O($N_s^2$). Further, each run of the CIM, emulated by integration of the CIM equations starting from a random initial state, is independent of any other run, allowing us to compute them in parallel. 

%% file: design.tex
\section{Design}
In this section, we discuss the design of our proposed \textit{Multi-Stage Delta Ising MIMO}~(MDI-MIMO). We describe the two key components of the MDI-MIMO algorithm: the degenerate delta Ising formulation for transforming the MIMO problem into an Ising problem, and a search space estimation methodology used to find an appropriately-reduced search space. We also discuss the flexibility in our algorithm and outline a few easy ways base stations can modify our proposal to match their performance-computation tradeoff requirements.
\label{sec:design}
\subsection{Degenerate Delta-Ising Formulation}
The key idea of our proposed degenerate Delta formulation is to assume a guess for $\mathbf{x_{ML}}$ in~(\ref{eq:realML-MIMO}), then optimize to find the best correction to the guess, rather than optimizing to find $\mathbf{x_{ML}}$ itself (unlike the existing state-of-the-art methods that use Quantum Annealing or CIMs).
 
Let us assume that we use the real-valued equivalent of the MMSE estimate~($\mathbf{x_m}$) as the guess: this step has low computational complexity and can be calculated from the MMSE solution using~(\ref{eq:realTransVec}). Given the solution variable $\mathbf{u}$ in the real-valued equivalent of the ML-MIMO problem~(\ref{eq:realML-MIMO}), define
\begin{equation}
    \mathbf{d} =  \mathbf{u} - \mathbf{x}_{m},
    \label{eq:d_define}
\end{equation}
where $\mathbf{d}$ can be interpreted as the correction to the MMSE solution. The goal then becomes the formulation of an Ising problem for computing the optimal $\mathbf{d}$, such that $\mathbf{x}_{m} + \mathbf{d}$ will be the solution to~(\ref{eq:realML-MIMO}). Note that any polynomial-time estimate of the MIMO problem, like Zero Forcing (ZF)~\cite{mm1}, MMSE, MMSE-SIC~\cite{optimalMMSEsic}, or Fixed complexity sphere decoder~(FSD)~\cite{fcsd}, can be used as the initial guess. We will discuss the impact of the choice of initial guess in Section~\ref{sec:initguess}.

For a QAM constellation, each element of $\mathbf{d}$ corresponds to the difference between the real and imaginary parts of two QAM symbols, and hence is an even-valued integer. The choice of the search space for $\mathbf{d}$ is a design parameter; and each element of $\mathbf{d}$ can take values in the set $\mathcal{D}_K = \{-2K,-2K+2,...,2K\}$, where $K$ is an appropriately chosen natural number (we will address the choice of $K$ in Section~\ref{sec:design:ssestimation}).  
The choice of the search space for $\mathbf{d}$ reflects the region around the MMSE estimate that is searched for the optimal solution and determines the resulting Ising formulation. Let $\bar{\mathbf{y}} = \mathbf{y}-\mathbf{H}\mathbf{x}_{m}$ be the residual received vector; then our new proposed formulation is obtained by substituting~(\ref{eq:d_define}) into~(\ref{eq:realML-MIMO}) to change the optimization variable to $\mathbf{d}$, and changing the search space to $\mathcal{D}_K^{2\cdot N_t}$:
\begin{equation}
    \mathbf{\hat{d}} = \arg \min_{\mathbf{d} \in \mathcal{D}_K^{2\cdot N_t}} ||\bar{\mathbf{y}}-\mathbf{H}\mathbf{d}||^2.
    \label{eq:diIsing}
\end{equation}
The next step is to express $\mathbf{d}$ using \textit{spin} variables which can only take values $-1$ or $1$. Note that any scalar $c \in \{-2,0,2\}$ can be expressed using two \textit{spin} variables $s_1,s_2 \in \{-1,1\}$ as $c = s_1 + s_2$. Similarly, if $c \in \{-4,-2,0,2,4\}$ then it can be expressed using four \textit{spin} variables as $c = s_1+s_2+s_3+s_4$. Applying this reasoning for the vector $\mathbf{d}$, we can express $\mathbf{d}$ as
\begin{equation}
    \mathbf{d} = \mathbf{T}\mathbf{s},
    \label{eq:T_transform}
\end{equation}
 where $\mathbf{T}$ denotes a transform matrix for $\mathcal{D}_K = \{-2K,-2K+2,...,2K\}$ where $K = 1,2...$, and $\mathbf{s}$ is a ($4KN_t\times 1$) spin vector (each element is $\pm1$). The transform matrix $\mathbf{T}$ is given by
 \begin{equation}
     \mathbf{T} = \mathbf{1}_{1\times 2K} \otimes \mathbf{I_{2 \cdot N_t}}.
     \label{def:transformMatrix}
 \end{equation}
where $\otimes$ is the Kronecker product.
We call this the \textit{degenerate delta Ising}~(dDI) formulation, as all spin vectors involved in the transform have the same coefficient. We get the equivalent Ising formulation of~(\ref{eq:realML-MIMO}) by substituting~(\ref{eq:T_transform}) in~(\ref{eq:diIsing}):
 \begin{equation}
     \hat{\mathbf{s}} = \arg \min_{\mathbf{s} \in \{-1,1\}^{2\cdot N_t}} -\mathbf{h}^T\mathbf{s} - \mathbf{s}^T\mathbf{J}\mathbf{s},
     \label{eq:IsingMIMO}
 \end{equation}
 \begin{equation}
     \begin{array}{lll}
       \mathbf{J}    &  = -\mathbf{T}^{T}\mathbf{H}^{T}\mathbf{HT}
       & = -\mathbf{1}_{2K \times 2K} \otimes \mathbf{H}^{T}\mathbf{H}\\
       \mathbf{h} &= \text{ }2\mathbf{T}^{T}\mathbf{H}^{T}\bar{\mathbf{y}} &= \text{ } \mathbf{1}_{2K\times 1} \otimes 2\mathbf{H}^{T}\bar{\mathbf{y}}.
     \end{array}
    \label{eq:IsingCoeff}
 \end{equation}
Note that the diagonal terms of $\mathbf{J}$ do not matter in the optimization process as they contribute a constant value to the objective function (given $\forall i, s_i^2 = 1$), and therefore, can be set to zero. We solve the obtained Ising problem (using an Ising machine) to get $\mathbf{\hat{s}}$, and then compute $\hat{\mathbf{d}} = \mathbf{T}\mathbf{\hat{s}}$ and $\mathbf{x} = \mathbf{x_m} + \mathbf{\hat{d}}$. Since some elements of $\mathbf{x}$ can be outside the QAM constellation, we round $\mathbf{x}$ to the nearest QAM symbol. Finally, we can get the maximum likelihood MIMO solution, $\mathbf{\Tilde{x}}$ in~(\ref{eq:ML-MIMO}), from $\mathbf{x}$ by inverting the transform described in~(\ref{eq:realTransVec}). 

The Ising problem obtained in~(\ref{eq:IsingMIMO}) has both linear and quadratic terms; however, as we saw in Section~\ref{sec:cim}, CIM requires an Ising problem with only quadratic terms. We use an auxiliary spin variable~\cite{ri-mimo} to convert all linear terms in~(\ref{eq:IsingMIMO}) to quadratic and define an auxiliary Ising problem:
 \begin{equation}
     (\bar{\mathbf{s}},\bar{s_a}) = \arg \min_{\mathbf{s} \in \{-1,1\}^{2N_t},s_a\in\{-1,1\}} -(\mathbf{h}^T\mathbf{s})s_a - \mathbf{s}^T\mathbf{J}\mathbf{s}.
     \label{eq:ising_aux}
 \end{equation}
 Note that, when $s_a = 1$, (\ref{eq:ising_aux})
is equivalent to (\ref{eq:IsingMIMO}). Therefore if the optimal solution of (\ref{eq:ising_aux}) has $\bar{s_a} = 1$, then it is also the optimal solution for (\ref{eq:IsingMIMO}). Further, given (\ref{eq:ising_aux}) is homogenous, if ($\bar{s_a} = -1,\bar{\mathbf{s}}$) is the optimal solution of (\ref{eq:ising_aux}), then ($\bar{s_a} = 1,-\bar{\mathbf{s}}$) is also the optimal solution, which implies $-\bar{\mathbf{s}}$ is the optimal solution of (\ref{eq:IsingMIMO}). Therefore, the solution of~(\ref{eq:IsingMIMO}) can be obtained from the solution of the auxiliary Ising problem~\cite{ri-mimo}  using the equation $\hat{\mathbf{s}} = \bar{\mathbf{s}}\times \bar{s_a}$. This allows us to solve~(\ref{eq:IsingMIMO}) on the CIM. 

We solve each problem on the amplitude heterogeneity correction-based CIM (described by~(\ref{eq:cim1}) and~(\ref{eq:cim2})). Borrowing the terminology from the Quantum Annealing literature, we refer to each run on the CIM as an \textit{anneal}. A standard practice is to run multiple anneals, for the same problem, to enhance the probability of finding the optimal solution. Hence, we run $N_a$ anneals for each problem instance, \textit{i.e.}, we solve each problem $N_a$ times on the CIM and generate $N_a$ candidate solutions. We compare the obtained candidate solutions and the initial guess and select the best solution based on the cost of the objective function in~(\ref{eq:realML-MIMO}).
\subsection{Search Space Estimation}
\label{sec:design:ssestimation}
The central idea of MDI-MIMO is to find the optimal correction to the polynomial-time guess to reach the ground state. However, if the polynomial-time guess is already correct or very close to the optimal solution, we can skip the optimization procedure completely. Further, the search space of the problem can be reduced by estimating the distance to the optimal solution. Therefore, having a mechanism to quantify the error in the initial guess can help us significantly reduce the computational load and appropriately determine the required search space. 

Consider the MDI-MIMO problem given by~(\ref{eq:diIsing}),
\begin{equation}
    \mathbf{\hat{d}} = \arg \min_{\mathbf{d} \in \mathcal{D}^{2\cdot N_t}} ||\bar{\mathbf{y}}-\mathbf{H}\mathbf{d}||^2.
\end{equation}
Note that $\mathbf{d} = 0$ corresponds to the initial guess. If a better solution exists, it implies
\begin{flalign}
    \exists \text{  }\mathbf{d}:& \text{ }||\bar{\mathbf{y}}-\mathbf{H}\mathbf{d}||^2 < ||\bar{\mathbf{y}}||^2\\
     \implies \exists \text{  }\mathbf{d}:& \text{ }\mathbf{d}^T(\mathbf{H}^T \mathbf{H})\mathbf{d} - 2(\mathbf{\bar{y}}^T\mathbf{H})\mathbf{d} < 0.
\end{flalign}
Since $(\mathbf{H}^T \mathbf{H})$ is a positive semi-definite matrix and has non-negative eigen values ($\lambda_1, \lambda_2 ... \lambda_{2\cdot N_T}$), by eigen-value decomposition
$\mathbf{H}^T \mathbf{H} = \mathbf{Q}^T\mathbf{\Sigma}\mathbf{Q}$, where $\mathbf{\Sigma} = diag(\lambda_1, \lambda_2 ... \lambda_{2\cdot N_T})$. Further, since $\mathbf{Q}$ is Unitary, $|| \mathbf{Qd}|| = || \mathbf{d}|| \geq K$, if $\mathbf{d}$ has at-least one element equal to $K$.
\begin{equation*}
    \forall  \mathbf{d}, \mathbf{d}^T(\mathbf{H}^T \mathbf{H})\mathbf{d} - 2(\mathbf{\bar{y}}^T\mathbf{H})\mathbf{d} > \lambda_{min}||\mathbf{Qd}||^2 - 2||\mathbf{\bar{y}}^T\mathbf{H}||||\mathbf{d}||,
\end{equation*}
which implies that
\begin{equation*}
      \forall  \mathbf{d}, \mathbf{d}^T(\mathbf{H}^T \mathbf{H})\mathbf{d} - 2(\mathbf{\bar{y}}^T\mathbf{H})\mathbf{d} > (\lambda_{min}||\mathbf{d}|| - 2||\mathbf{\bar{y}}^T\mathbf{H}||)\cdot||\mathbf{d}||.
\end{equation*}
Therefore, $ \lambda_{min}K - 2||\mathbf{\bar{y}}^T\mathbf{H}|| > 0 \implies$ no solution with at least one element equal to K exists.

Therefore, for a better solution (with at least one element equal to $K$) to exist, the necessary condition is given by 
\begin{equation}
    K < \dfrac{2||\mathbf{\bar{y}}^T\mathbf{H}||}{\lambda_{min}}.
\end{equation}

Let 2$\omega$ be the highest difference between the I/Q components of symbols drawn from some $M$-QAM modulation. For instance, $2\omega = 2$ for 4-QAM, $2\omega = 4$ for 16-QAM, and $2\omega = 14$ for 64-QAM. 
Since each element $\hat{d}_i$ of $\hat{\mathbf{d}}$ in~(\ref{eq:diIsing}) can only take even values, the appropriate search space can be determined as follows:
if $\Gamma$ is defined as $\Gamma = \dfrac{2||\mathbf{\bar{y}}^T\mathbf{H}||}{\lambda_{min}}$, then the search space is given by
\begin{equation}
    \mathcal{D_{\omega}} = \{-2\omega,-2\omega+2,...2\omega\}, \text{ if }  2\omega \leq \Gamma
\end{equation}
\textit{i.e.} when the predicted search space is larger than the maximum possible search space. Otherwise, the appropriate search space is given by:
\begin{equation*}
    \mathcal{D}_R = \{-2R,-2R+2,...2R\}, 
\end{equation*}
\begin{equation}
    \text{ s.t    }  0 \leq 2R < 2\omega \text{    and    }\Gamma - 2 < 2R \leq \Gamma,
\end{equation}
However, using $\lambda_{min}$ leads to an extremely conservative metric, and we can relax this exact metric to a heuristic which uses $\lambda_{mean}$ or $\lambda_{max}$ instead of $\lambda_{min}$, \textit{i.e.}, $\Gamma = \frac{2||\mathbf{\bar{y}}^T\mathbf{H}||}{\lambda_{mean}}$ or $\Gamma = \frac{2||\mathbf{\bar{y}}^T\mathbf{H}||}{\lambda_{max}}$. We will show in Section~\ref{sec:eval:ssestimation} that this procedure has good accuracy in predicting the search space and significantly reduces the computation load while having an acceptable impact on the performance (especially in mid-high SNR scenarios).
\begin{figure}[h!]
    \centering
    \includegraphics[width=1\linewidth]{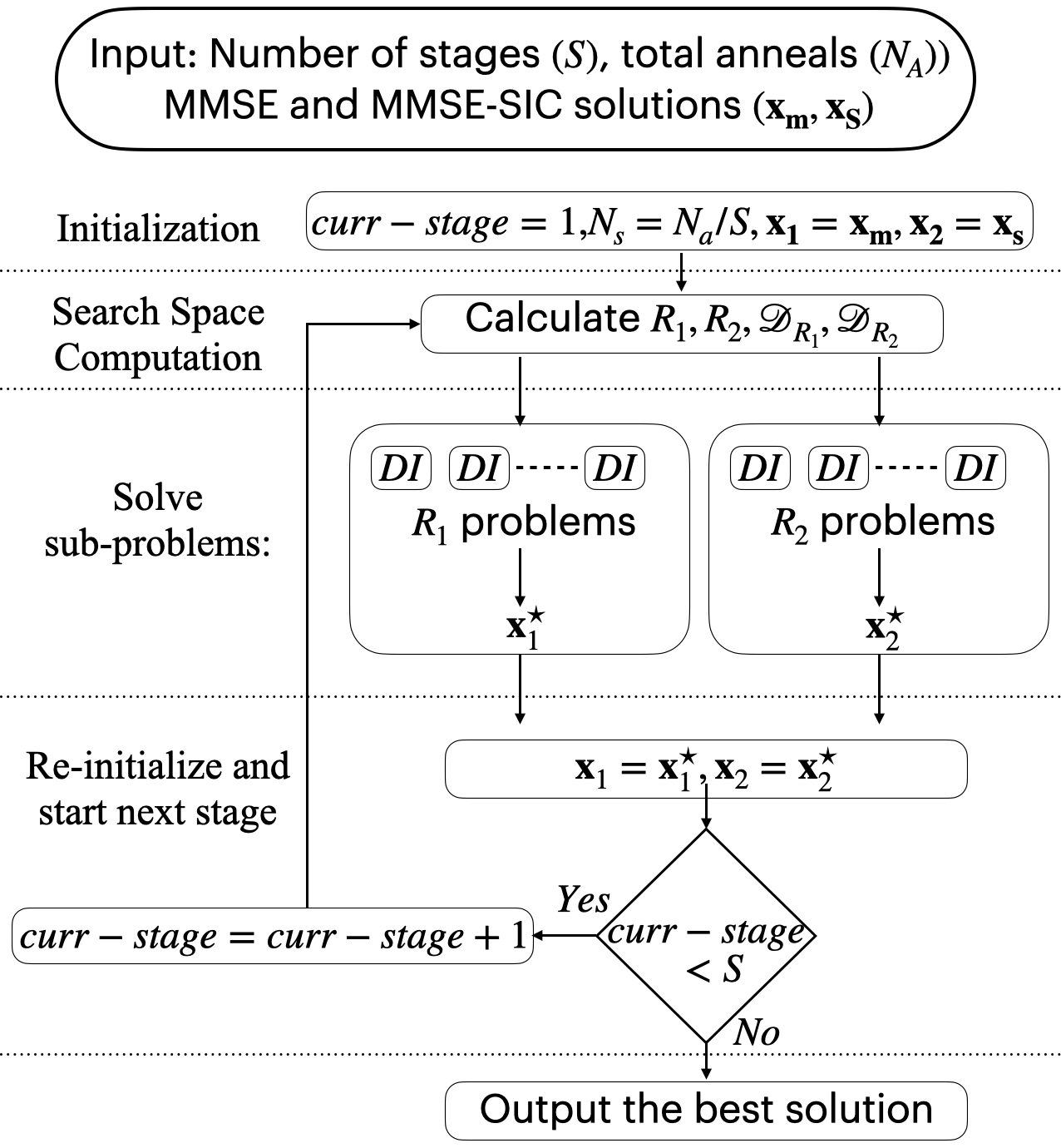}
    \caption{MDI-MIMO: flow diagram illustrating various steps in the algorithm.} 
    \label{fig:diFlow}
\end{figure}
\begin{figure*}[h!]
    \centering
    \includegraphics[width=1\textwidth]{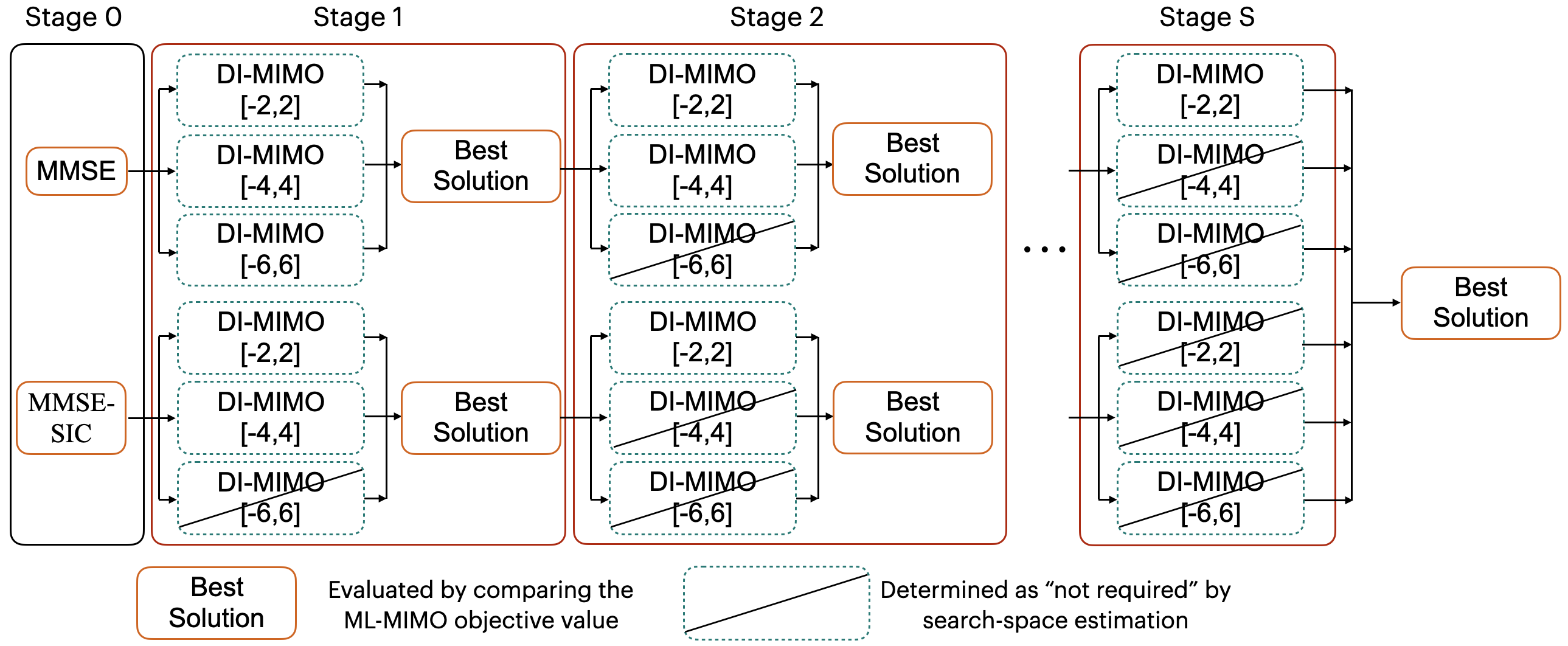}
    \caption{MDI-MIMO: example execution for 16-QAM illustrating that MDI-MIMO adaptively adjusts the search space and improves the starting guess.} 
    \label{fig:diMimoIntro}
\end{figure*}
\subsection{Multi-Stage Delta Ising-MIMO~(MDI-MIMO)}
\label{sec:mdimimodesign}
In this section, we describe the proposed Multi-Stage-Delta-Ising-based MIMO detection~(MDI-MIMO) algorithm. Given a MIMO instance and a budget of $N_a$ total anneals, the MDI-MIMO algorithm~(that allows $S$ stages and explores at most $R_{max}$ neighbors) is as follows:
\begin{enumerate}
    \item Calculate the number of anneals for each stage  $N_s = N_a/S$ (by splitting total anneals equally among all stages).
    \item Set the index of the initial stage, $current-stage = 1$.
    \item Calculate the MMSE estimate~($\mathbf{x}_m$) and MMSE-SIC~($\mathbf{x}_s$) estimate of the MIMO problem. 
    \item set $\mathbf{x}_1 = \mathbf{x}_m$ and $\mathbf{x}_2 = \mathbf{x}_s$. These will serve as the initial guesses for the dDI formulation.
    \item Compute the appropriate search space around $\mathbf{x}_1$. Let this be $\mathcal{D}_{R_1}$, such that $R_1 \leq R_{max}$~(if the computed $R_1$  is larger than $R_{max}$ we set it equal to $R_{max}$). 
    \item Repeat previous step, for $\mathbf{x}_2$ and compute $R_2$.
    \item Next we run each of the ($R_1$ + $R_2$)  instances, computed in the last two steps with $N_s/(R_1 + R_2)$ anneals each. The instances correspond to initial guess $\mathbf{x}_1$ and search spaces ($\mathcal{D}_1,...\mathcal{D}_{R_1}$), initial guess  $\mathbf{x}_2$ and search spaces ($\mathcal{D}_1,...\mathcal{D}_{R_2}$).
    \item From the $R_1$ and $R_2$ instances corresponding to $\mathbf{x_1}$ and $\mathbf{x_2}$ respectively, select the best solutions $\mathbf{x^\star_1}$ and $\mathbf{x^\star_2}$.
    \item If $current-stage = S$ go to Step~10, otherwise set $\mathbf{x}_1 = \mathbf{x^\star_1}$, $\mathbf{x}_2 = \mathbf{x^\star_2}$, increment $current-stage$ by 1, and go to Step~5.
    \item Output the better solution out of $\mathbf{x^\star_1}$ and $\mathbf{x^\star_2}$ as the solution. 
\end{enumerate}
The goal of Step~5 is to create multiple parallel instances of DI-MIMO in each symmetric subset of $\mathcal{D}_{R_1}$. Therefore if $\mathcal{D}_{R_1} = \{-2R_1,...2R_1\}$, then we would create $R_1$ parallel instances of DI-MIMO in search spaces: $\{-2,2\},\{-4,4\}...\{-2R_1,2R_1\}$. We will justify this design choice in Section~\ref{sec:multipleSearch}. Fig.~\ref{fig:diFlow} is a flow diagram summarizing MDI-MIMO and Fig.~\ref{fig:diMimoIntro} provides an example execution of MDI-MIMO for 16-QAM modulation.

\subsection{Generalizing MDI-MIMO}
\label{sec:generalizeMDI}
Our proposed multi-stage algorithm aims to improve any approximate solution for MIMO using Ising machines. In this paper, the proposed MDI-MIMO algorithm starts with two approximate solutions for the MIMO problem: MMSE and MMSE-SIC. Given their low complexity and availability of efficient FPGA/ASIC-based implementations, they serve as good initial guesses. However, the multi-stage approach adopted in MDI-MIMO can be easily extended to use any number of starting guesses. The multi-stage search corresponding to different guesses operates independently, and therefore it is very easy for the base station to modify the number of starting guesses based on computational load and QoS requirements. If the base station has idle compute resources, it can choose to improve the performance further by executing the MDI-MIMO approach with more than two starting guesses (let's say MMSE, MMSE-SIC, and FSD~\cite{fcsd}). If the base station is experiencing a compute resource shortage, the base station can choose to simplify the algorithm and only use MMSE as the starting guess. We will see later (Section~\ref{sec:initguess}) that even this simplified version can provide significant gains. 
\subsection{Computational complexity of MDI-MIMO}
In Section~\ref{sec:ising_conv}, we showed that the complexity of a single anneal is O($N_s^2$), where $N_s$ are the number of spins in the problem. The largest sub-problem of MDI-MIMO, corresponding to search in $\{-2R_{max},-2R_{max}+2,...2R_{max}\}$, will require $2N_t\cdot2R_{max}$ spin variables and one spin variable to convert all linear terms to be quadratic. Therefore, for any MDI-MIMO anneal $N_s \leq (4N_tR_{max} + 1)$. For an M-QAM constellation, $R_{max} \leq \lceil\sqrt{M}\rceil$, which implies $N_s \leq (4N_t\lceil\sqrt{M}\rceil + 1)$. Therefore, the complexity of each MDI-MIMO anneal is O($MN_t^2$) and the complexity for $N_a$ anneals is given by O($N_aMN_t^2$). Further, the complexity of calculating Ising coefficients is dominated by calculating $\mathbf{H}^{T}\mathbf{H}$ and is given by O($N_t^2N_r$). In order to determine the quality of states produced by CIM, the ML objective function needs to be evaluated, which has a complexity of $O(N_aN_rN_t)$. Finally, since MDI-MIMO needs one or more initial guesses, the total complexity of MDI-MIMO can be given as:
\begin{equation*}
    O(N_aMN_t^2 + N_t^2N_r + N_aN_rN_t)+\text{comp. of initial guess}.
\end{equation*}

Note that, MDI-MIMO's contribution to the complexity is a polynomial in the number of anneals ($N_a$), number of antennas at the base station ($N_r$), and number of users ($N_t$). Therefore, it is necessary to use initial guesses with polynomial complexity that are quick to compute (\textit{e.g.}, MMSE, ZF, MMSE-SIC) and not ones with exponential-complexity (\textit{e.g.}, K-Best tree search~\cite{kbest}). As discussed in Section~\ref{sec:mdimimodesign}, in the proposed design we start with two initial guesses (MMSE and MMSE-SIC) and split the computational resources. 

%% file: eval.tex
\section{Evaluation}
\label{sec:eval}
In this section, we evaluate our proposed algorithm for large and massive MIMO systems. We perform several benchmarking experiments to optimize the free parameters of our algorithm and justify our design choices. We further compute the BER and spectral efficiency of our methods for several MIMO systems with a large number of antennas and users. Finally, we demonstrate the performance of our methods in a realistic 3GPP-compliant LTE scenario. 
\begin{figure*}[h]
  \centering
  \includegraphics[width=1\textwidth]{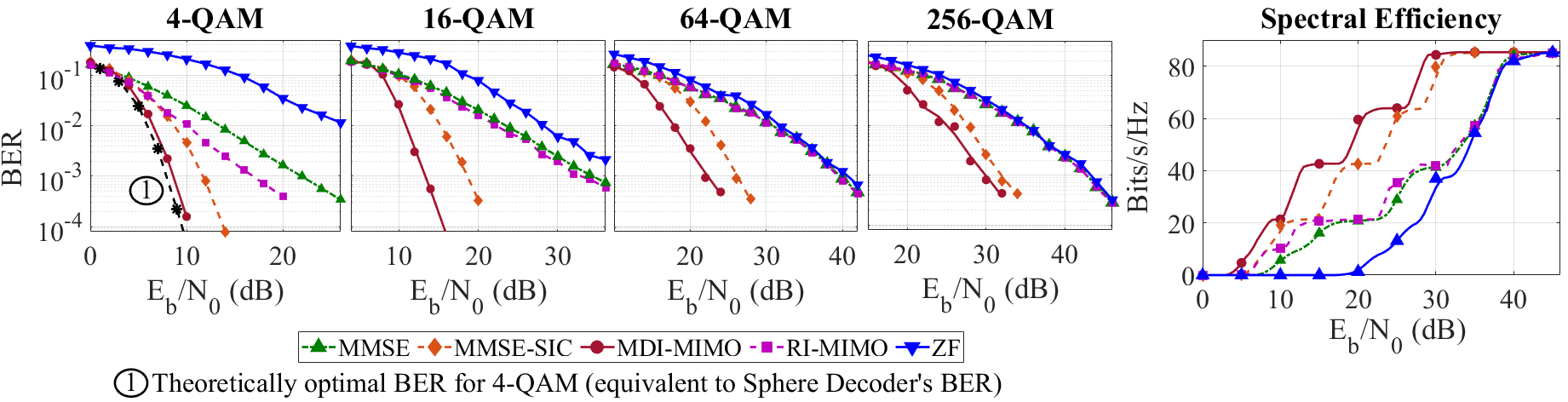}
  \caption{BER and spectral efficiency performance of ($16$ users $\times$ $16$ antennas) Large MIMO: demonstrating that MDI-MIMO significantly outperforms MMSE, MMSE-SIC, ZF, and RI-MIMO and provides much higher spectral efficiency.} 
  \label{fig:ber16}
\end{figure*} 
\begin{figure*}[h]
  \centering
  \includegraphics[width=1\textwidth]{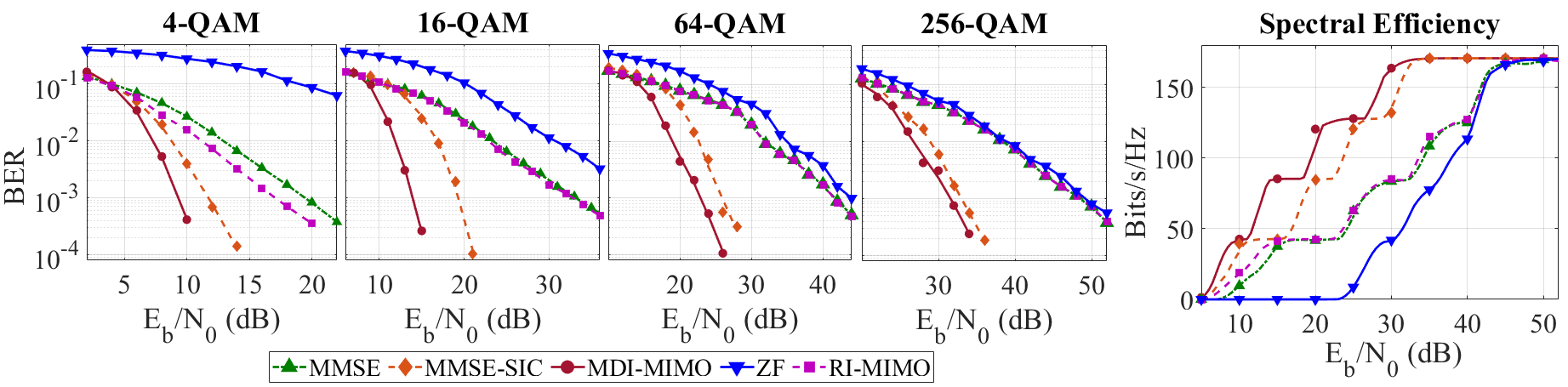}
  \caption{BER and spectral efficiency performance of ($32$ users $\times$ $32$ antennas) Large MIMO: demonstrating that MDI-MIMO significantly outperforms  ZF, MMSE, MMSE-SIC, and RI-MIMO and provides much higher spectral efficiency.} 
  \label{fig:ber32}
\end{figure*}
\begin{figure*}[h]
  \centering
  \includegraphics[width=1\textwidth]{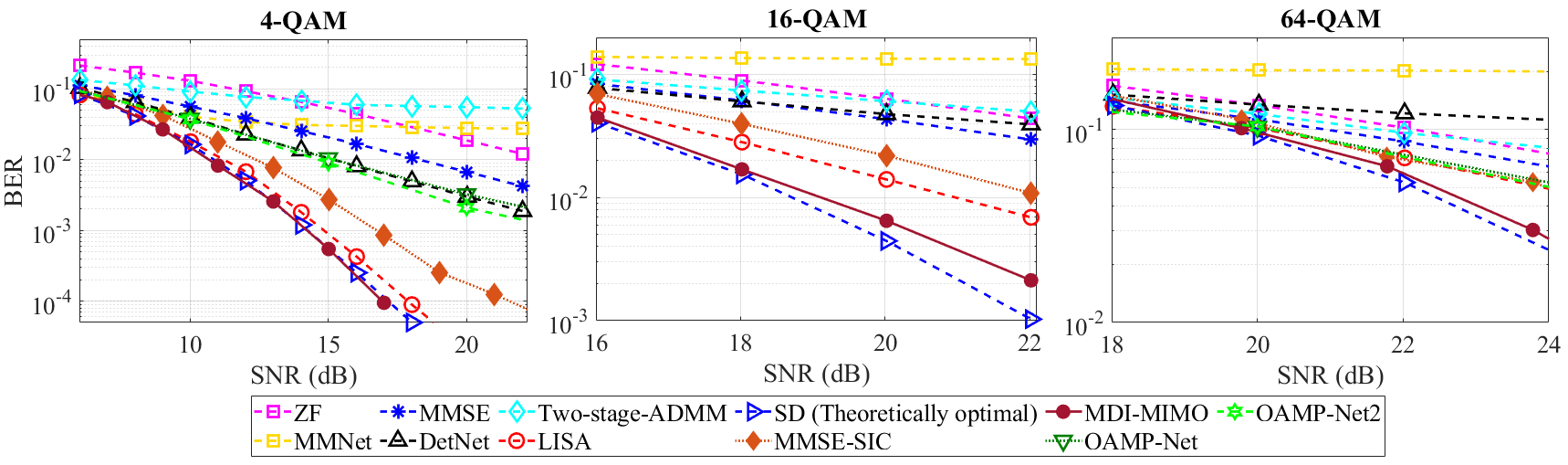}
  \caption{Comparing BER performance of MDI-MIMO to machine learning-based methods for ($4$ users $\times$ $4$ antennas) MIMO: we see that our proposed MDI-MIMO can outperform the machine learning algorithms and achieve near-optimal performance for ($4$ users $\times$ $4$ antennas) MIMO.\\\textit{*data unavailable for 16-QAM OAMP-Net and OAMP-Net2}} 
  \label{fig:ber_ml}
\end{figure*}
\subsection{Evaluation Setup}
Our evaluation setup simulates an uplink $N_t\times N_r$ MIMO system which has $N_t$ users (with one transmit antenna each) and $N_r$ receive antennas at the base station ($N_r \geq N_t$). We assume a slow-fading channel and channel instances are assumed to follow the Rayleigh fading model. 
\subsubsection{Implementation of the CIM Simulator}
In this paper, we implement an Ising solver by simulating the time evolution of the amplitude heterogeneity correction-based bi-stable system described by~(\ref{eq:cim1}) and~(\ref{eq:cim2}). Our simulator performs numerical integration using the Range-Kutta method~(RK4)~\cite{rk4} for 512 time~steps and $dt = 0.005$. Note that the step size required for convergence of numerical integration will become increasingly smaller as the problem size increases. For the scenarios simulated in this paper, $dt = 0.005$ was found to be sufficient.  The initial values $x_i$ are i.i.d $\mathcal{N}(0,0.001)$ and $e_i$ are initialized using a folded $\mathcal{N}(0,0.001)$ distribution. We set $p = 0.98$, $\beta = 1$, $a = 2$ and $\gamma = 1000/(512\cdot0.005)$. These parameters were empirically selected, based on trial-and-error experiments, such that the system can achieve a steady state and attain good performance. Note that performance can be further improved by optimally selecting these parameters, and we plan to address this in our future work. If any iteration fails to converge in 512 integration steps, then the spin output $\mathbf{s}$ is such that $\mathbf{T}\mathbf{s} = \mathbf{0}$~(where $\mathbf{T}$ is the transform matrix defined by~(\ref{def:transformMatrix})). Additionally, we also consider the intermediate states of the CIM as candidate solutions as they are readily available to the solver after each integration step. 
\subsubsection{Evaluation Metrics}
We primarily use two evaluation metrics: BER and spectral efficiency. The BER is computed as the mean BER of all users. We compare our methods against MMSE-SIC with optimal ordering~\cite{optimalMMSEsic}, MMSE detector, and RI-MIMO~\cite{ri-mimo} in both large and massive MIMO scenarios. Spectral efficiency computations are based on convolutional coding with code-rates $\frac{1}{3}$, $\frac{1}{2}$, $\frac{2}{3}$, and an oracle Adaptive Modulation and Coding~(AMC) module that selects the best modulation and code-rate based on SNR. 
\begin{figure*}[h]
  \centering
  \includegraphics[width=1\textwidth]{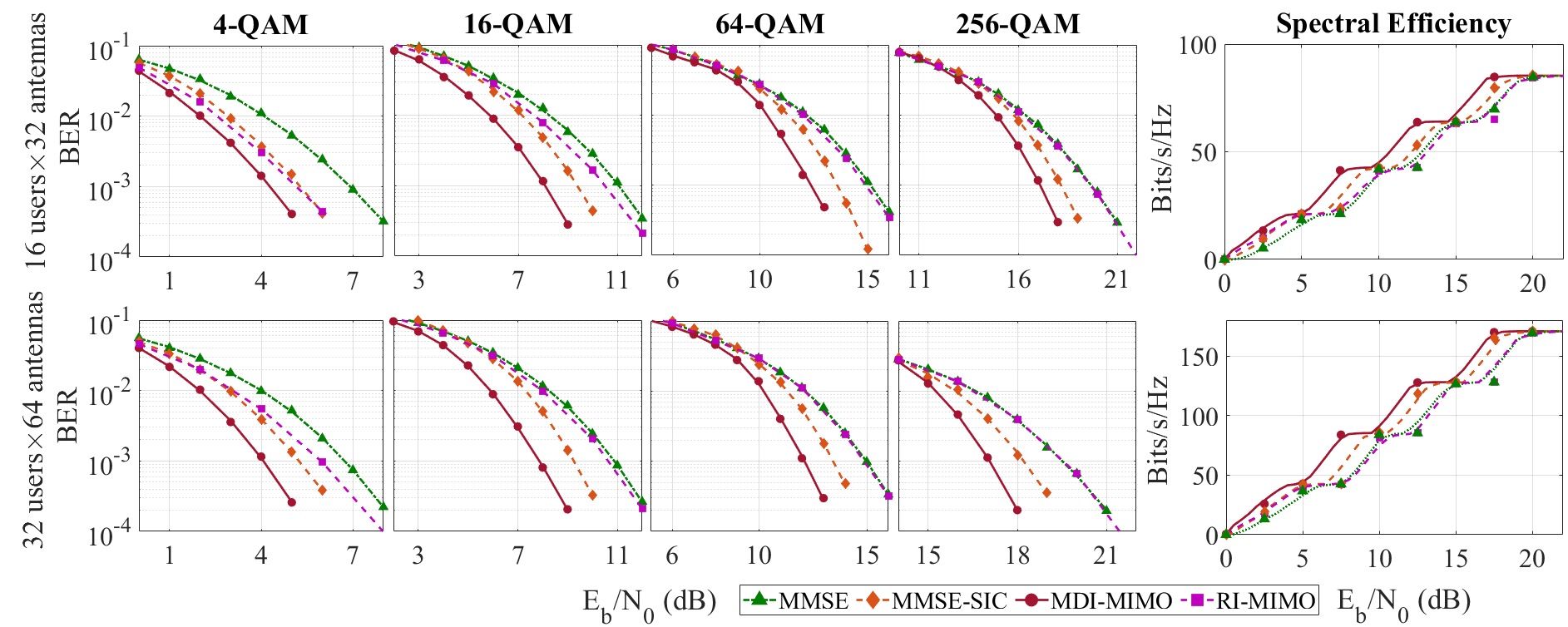}
  \caption{BER and spectral efficiency performance of ($16$ users $\times$ $32$ antennas) and ($32$ users $\times$ $64$ antennas) Massive MIMO: demonstrating that MDI-MIMO significantly outperforms MMSE, MMSE-SIC, and RI-MIMO and provides much higher spectral efficiency.} 
  \label{fig:bermm}
\end{figure*}
\begin{figure*}[h]
  \centering
  \includegraphics[width=1\textwidth]{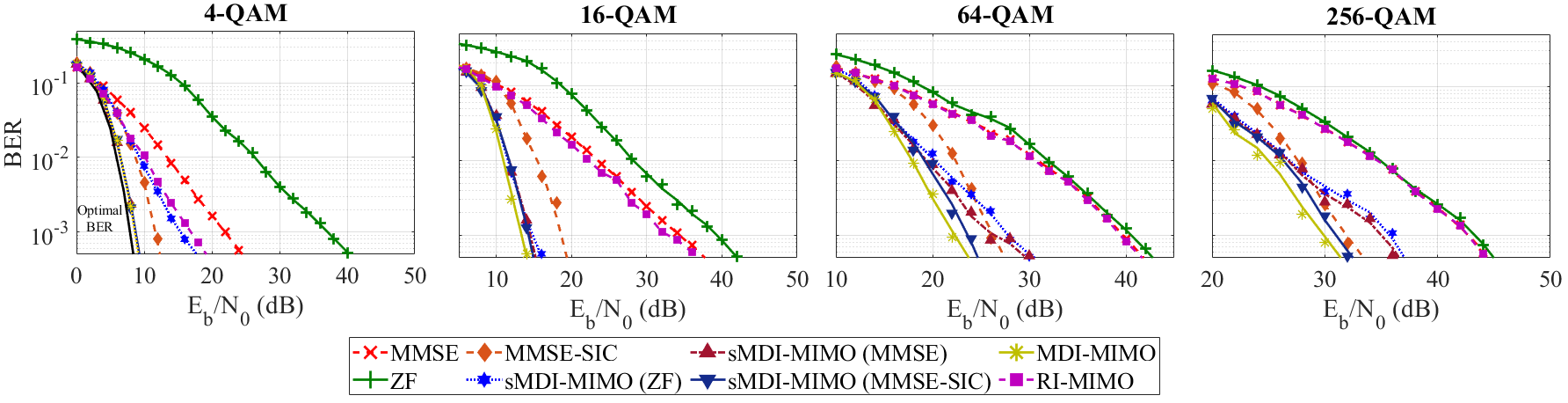}
  \caption{MDI-MIMO BER performance of ($16$ users $\times$ $16$ antennas) Large MIMO: demonstrating the variation in the BER performance with different starting guesses for MDI-MIMO.} 
  \label{fig:berinit}
\end{figure*}
\begin{table}[h!]
\caption{Experimental parameters for MDI-MIMO defined using a 4-tuple ($N_s$, heuristic, $R_{max}$, $S$). Note that, we choose $\lambda_{max}$ heuristic for ($32$ users $\times$ $32$ antennas, 256-QAM) to reduce the execution time (see Section~\ref{sec:eval:ssestimation}), given the problem size is much higher at $R_{max} = 8$.}
\centering
\begin{tabular}{ccc} 
\toprule
& \textbf{4-QAM} & \textbf{16-QAM}\\
& \textbf{DI-MIMO} & \textbf{MDI-MIMO}\\
\toprule
\textbf{$\mathbf{16}$ users $\mathbf{\times}$ $\mathbf{16}$ antennas} &$(64,\lambda_{mean},1,1)$ &$(64,\lambda_{mean},3,8)$\\
\textbf{$\mathbf{32}$ users $\mathbf{\times}$ $\mathbf{32}$ antennas}   &$(64,\lambda_{mean},1,1)$ &$(128,\lambda_{mean},3,8)$\\
\textbf{$\mathbf{16}$ users $\mathbf{\times}$ $\mathbf{32}$ antennas}   &$(64,\lambda_{mean},1,1)$ &$(8,\lambda_{mean},3,8)$\\
\textbf{$\mathbf{32}$ users $\mathbf{\times}$ $\mathbf{64}$ antennas}   &$(64,\lambda_{mean},1,1)$ &$(8,\lambda_{mean},3,8)$\\
\end{tabular}

\begin{tabular}{ccc} 
\toprule
& \textbf{64-QAM} & \textbf{256-QAM}\\
& \textbf{DI-MIMO} & \textbf{MDI-MIMO}\\
\toprule
\textbf{$\mathbf{16}$ users $\mathbf{\times}$ $\mathbf{16}$ antennas} &$(64,\lambda_{mean},4,8)$ &$(64,\lambda_{mean},8,8)$\\
\textbf{$\mathbf{32}$ users $\mathbf{\times}$ $\mathbf{32}$ antennas}  &$(128,\lambda_{mean},4,8)$ &$(128,\lambda_{max},8,8)$\\ 
\textbf{$\mathbf{16}$ users $\mathbf{\times}$ $\mathbf{32}$ antennas}  &$(8,\lambda_{mean},4,8)$ &$(8,\lambda_{mean},8,8)$\\
\textbf{$\mathbf{32}$ users $\mathbf{\times}$ $\mathbf{64}$ antennas}   &$(8,\lambda_{mean},4,8)$ &$(8,\lambda_{mean},8,8)$\\
\bottomrule
\end{tabular}
\label{tbl:params}
\end{table}
\subsection{Performance Analysis}
In this section, we emulate the end-to-end performance of several large MIMO and massive MIMO systems. We assume an oracle Adaptive Modulation and Coding~(AMC) module, which selects the best modulation and coding scheme~(among the available options) for every SNR. Such an oracle AMC allows us to approximate the spectral efficiency of the system empirically. We allow our system to use, 4-, 16-, 64-, and 256-QAM modulation and convolutional coding with $\frac{1}{3}$, $\frac{1}{2}$, and $\frac{2}{3}$. For 4-QAM, given that the problem is significantly simpler, MDI-MIMO only uses MMSE as the starting guess and has only one stage. For higher modulations, it uses both MMSE and MMSE-SIC as the starting guesses. The parameters of MDI-MIMO can be described using a 4-tuple ($N_s$, heuristic, $R_{max}$, $S$). Our simulation parameters are reported in Table~\ref{tbl:params}.

In Fig.~\ref{fig:ber16}, we simulate a ($16$ users $\times$ $16$ antennas) MIMO system and compare the performance of MMSE, MMSE-SIC, ZF, RI-MIMO~\cite{ri-mimo}, and our proposed MDI-MIMO. Note that, as computing the theoretical optimal (Sphere Decoder~\cite{sphereIeee}) has exponential complexity, it is only feasible to compute it for 4-QAM modulation. We see that for 4-QAM modulation, MDI-MIMO is near-optimal and significantly better than all other tested algorithms. For 16-QAM and higher modulations, MDI-MIMO significantly outperforms all other methods, and we see that the multi-stage approach of MDI-MIMO, which adaptively adjusts the search radius and improves the initial guess, can provide significant performance improvements over all other tested algorithms in every test scenario. Next, we simulate a ($32$ users $\times$ $32$ antennas) MIMO system in Fig.~\ref{fig:ber32} and ($16$ users $\times$ $32$ antennas), ($32$ users $\times$ $64$ antennas) massive MIMO systems in Fig.~\ref{fig:bermm}, and see that MDI-MIMO can provide significant performance improvements over all other tested algorithms. In these experiments, we focus on MIMO systems with a large number of users and antennas for which computing the theoretically optimal performance is computationally infeasible. Further, as noted before, these scenarios are not tackled by machine learning/AMP-based approaches. Therefore, next, we look at a smaller ($4$ users $\times$ $4$ antennas) scenario and compare the performance to the theoretically optimal Sphere decoder (SD) and several machine learning-based/AMP-based approaches. The parameters of MDI-MIMO are selected to be the same as in the ($16$ users $\times$ $16$ antennas) scenario.   

We compare MDI-MIMO to LISA~\cite{lisa}, MMNet~\cite{mmnet}, DetNet~\cite{detnet}, and Two-stage-ADMM~\cite{tsadmm}. Unlike LISA~\cite{lisa}, which plots BER vs the ratio to transmit power to noise power, we plot BER vs SNR (ratio of received power to noise power), and therefore, the x-axis of our results and results of LISA~\cite{lisa} mismatch. However, this does not impact the relative BER trends as it merely shifts all the curves equally along the x-axis. We also compare against two AMP-inspired ML algorithms, OAMP-Net~\cite{oampnet}, and OAMP-Net2~\cite{mlModeldriven}. In Fig.~\ref{fig:ber_ml}, we see that MDI-MIMO significantly outperforms the AMP/ML-based methodologies and provides near-optimal performance (close to the theoretically optimal SD) for all tested modulation schemes. 

The computational complexity of these algorithms can be summarized in Table~\ref{tbl:comp}. We see that, for Large MIMO scenarios ($N_t \approx N_r$), the complexity of the tested algorithms and MDI-MIMO is similar, however as shown before, MDI-MIMO vastly outperforms these algorithms. 
\begin{table}[h!]
\caption{Computational complexity of MDI-MIMO and other tested machine-learning algorithms}
\centering
\begin{tabular}{cc} 
\toprule
\textbf{MDI-MIMO} & O$(N_aMN_t^2 + N_t^2N_r + N_aN_rN_t)+\text{initial guess}$\\
LISA & O($N_t^2N_r + N_td_h^2 + N_t^2d_h$)~\cite{lisa}\\
DetNet & O($N_rN_t^2 + d_ZN_t + d_Zd_V$)~\cite{lisa}\\
MMNet & O($N_rN_t^2$)~\cite{lisa} \\
OAMP-Net & O($N_r^3$)~\cite{lisa}\\
OAMP-Net2 & O($TN_t^3$)~\cite{mlModeldriven}\\
\toprule
&*$d_h, d_Z, d_V, T$ are parameters of respective algorithms\\
\end{tabular}
\label{tbl:comp}
\end{table}

\begin{figure}[ht]
  \centering
  \includegraphics[width=\linewidth]{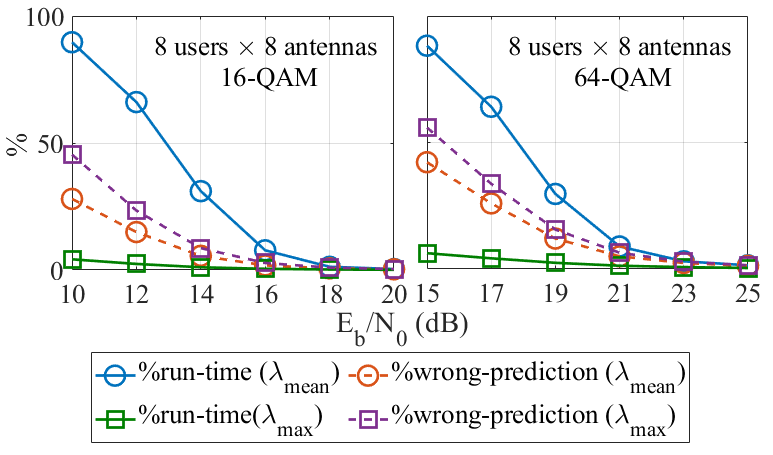}
  \caption{Analysis of search space estimation heuristics for ($8$ users $\times$ $8$ antennas) MIMO system with 16- and 64-QAM modulation: demonstrating a significant reduction in the complexity with low prediction error, especially at mid-high SNR.}
  \label{fig:pred12}
\end{figure}
\subsection{Impact of Initial Guess}
\label{sec:initguess}
In Section~\ref{sec:generalizeMDI}, we argued that given the modular design of MDI-MIMO, the base station can vary the number of initial guesses used by MDI-MIMO depending upon compute load and QoS requirements. The key idea is to use simple/fast well-known classical MIMO detection algorithms to generate the initial guess for MDI-MIMO. These guesses must be quick to compute and should not add significant computational load on the base station. For example, ZF and MMSE, being simple linear MIMO detectors, are suitable candidates, however algorithms like FSD~\cite{fcsd} (with high depth) which achieve good BER performance at the cost of large compute time, are not feasible initial guesses. To understand the impact of initial guess on performance, in this section,  we will look at the BER performance of different variants of our proposed methodology:
\begin{itemize}
    \item MDI-MIMO: which starts with MMSE and MMSE-SIC as the two initial guesses. 
    \item Simplified MDI-MIMO (sMDI-MIMO): which starts with a single initial guess. We explore ZF, MMSE, or MMSE-SIC as the initial guess for sMDI-MIMO. 
\end{itemize}
Note that for each method we provide the same amount of compute resources (for MDI-MIMO they are split between the two initial guesses). From Fig.~\ref{fig:berinit}, we can see that, for 4-QAM modulation, using MMSE or MMSE-SIC as initial guesses (individually or together) leads to similar BER performance which is close to the theoretical optimal. However, using ZF as the initial guess leads to a deterioration of BER performance. This can be attributed to the poor quality of ZF as an estimate of the MIMO problem (as seen from the massive difference between BER attained by MMSE and ZF). Note that for 4-QAM, MDI-MIMO only uses a single stage (as noted in Table~\ref{tbl:params}) and cannot improve the quality of the initial guess.

For 16-QAM, the BER performance is agnostic to the choice of initial guesses despite large differences in their quality. For 16-QAM and higher, our proposed methods use a multi-stage approach,  where at the end of each stage, the initial guess is changed to the best solution found until that point. Therefore, the quality of the initial guess is subsequently improved by the algorithm making it resilient to differences in the quality of the initial choice.  

For 64-QAM and 256-QAM, the BER is agnostic to the choice of the initial guess for low/mid-SNR scenarios and only differs for high-SNR scenarios. For high-SNR scenarios, the BER performance improves with the quality of the initial guess. The best performance is however achieved when both MMSE and MMSE-SIC are used as initial guesses (and splitting the compute resources). 
    
Based on these observations, our proposed design utilizes both MMSE and MMSE-SIC as the initial guesses (as discussed in Section~\ref{sec:mdimimodesign}). However, the proposed methodology is highly flexible and allows base stations to easily switch between different types/number of initial guesses depending upon the QoS requirements and SNR.

\subsection{Analysis of search space estimation}
\label{sec:eval:ssestimation}
In Section~\ref{sec:design:ssestimation}, we proposed a heuristic method to estimate the maximum required search space size for a problem instance. In this section, we will evaluate the error and complexity reduction of the proposed approaches. 
We simulate an ($8$ users $\times$ $8$ antennas) MIMO system, with MMSE-SIC as the initial guess for the dDI formulation, and evaluate the following:
\begin{enumerate}
  \item \%run-time =  fraction of instances required to be solved on the CIM, \textit{i.e.}, predicted search space size is greater than 0. 
  \item\%wrong-prediction = fraction of instances for which the optimal solution lies outside the estimated search space. 
\end{enumerate}
Further, we evaluate two scenarios: use of maximum~($\lambda_{max}$) or mean~($\lambda_{mean}$) eigenvalue (of the real-equivalent channel) for search space estimation (as described in Section~\ref{sec:design:ssestimation}). We observe in Fig.~\ref{fig:pred12}, that both methods provide a significant reduction in the number of instances for which the Ising machine needs to be used, hence, drastically reducing the computational load on the Ising machine. The prediction error can be high at low SNR, it rapidly reduces and becomes very low for mid and high-SNR scenarios. We also note that using $\lambda_{max}$ (compared to $\lambda_{mean}$) leads to a lower run-time at the cost of higher prediction error. In this paper, we use $\lambda_{mean}$ for our evaluation due to its lower prediction error, unless specified otherwise. 
\subsection{Advantages of degenerate Delta Ising representation}
\label{sec:eval:rep}
In this paper, we proposed the degenerate Delta Ising formulation (dDI), which is an improvement over our previously proposed Delta Ising formulation (DI)~\cite{di-mimo}. While both these representations are equivalent for search space $\mathcal{D}_1 =\{-2,0,2\}$ (which corresponds to searching among the nearest neighbors of the initial guess), they differ for larger search spaces and the proposed degenerate delta-Ising formulation~(dDI) performs much better. For example, the transform matrix (corresponding to (\ref{eq:T_transform})) for $\mathcal{D}_2$ is given by:
\begin{equation*}
\begin{array}{ll}
     \textit{DI~\cite{di-mimo}:}&  \mathbf{T} = [2\mathbf{I}_{2\cdot N_t} \text{ }\mathbf{I}_{2\cdot N_t} \text{ }\mathbf{I}_{2\cdot N_t}]\\
    \textit{dDI (PROPOSED):} & \mathbf{T} = [\mathbf{I}_{2\cdot N_t} \text{ }\mathbf{I}_{2\cdot N_t}  \text{ }\mathbf{I}_{2\cdot N_t} \text{ }\mathbf{I}_{2\cdot N_t}]\\
\end{array}
 \label{eq:T_def}
\end{equation*}

\begin{figure*}[ht]
  \centering
  \includegraphics[width=1\linewidth]{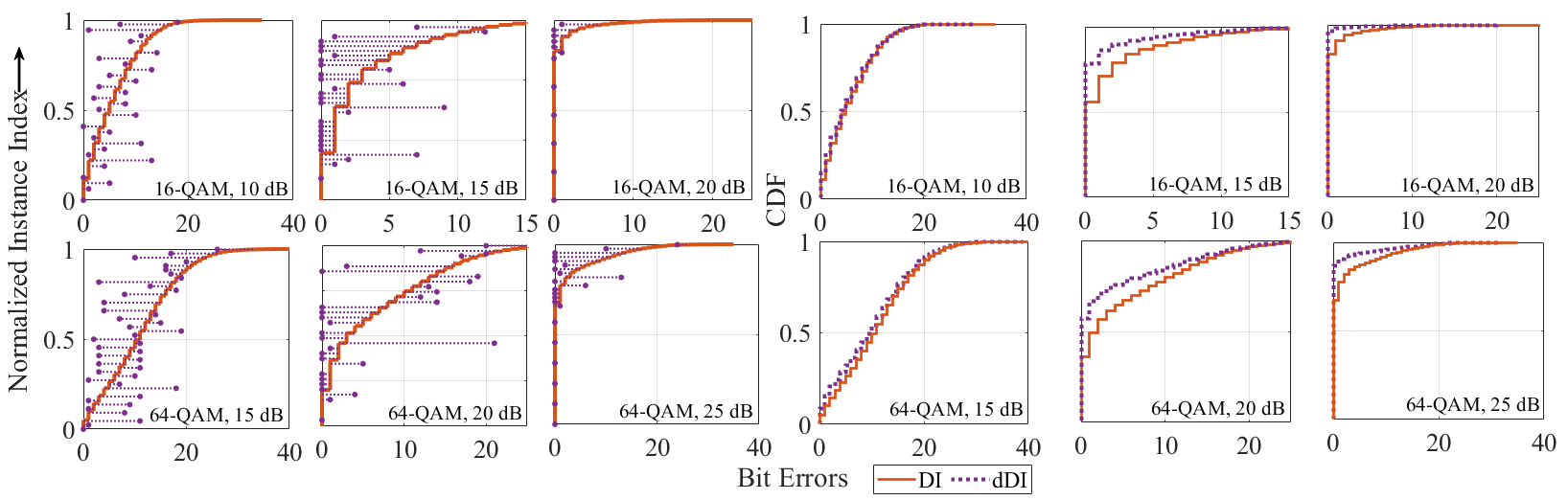}
  \caption{Coupled-plot demonstrating bit errors for previously proposed DI and the improved dDI formulation for ($16$ users $\times$ $16$ antennas) MIMO, 16- and 64-QAM systems: we see that dDI formulation leads to a much better overall probability of finding the ground state. Note that the instances not shown in the coupled-plots~(Left) have zero bit errors with both DI and dDI formulation.} 
  \label{fig:rep12}
\end{figure*}

In Fig.~\ref{fig:rep12}\footnote{In this paper, we often use \textit{coupled-plots} to describe several evaluation results. A coupled-plot between two entities (measured by same units) $E_1$ and $E_2$, where data corresponding to $E_1$ is in sorted ascending order, involves plotting $E_1$ on the x-axis and the corresponding instance index on the y-axis (this is equivalent to a CDF of $E_1$ multiplied by the total number of instances). For each data point (representing an instance of $E_1$), the corresponding value of $E_2$ is indicated using a dotted line. For readability of plots, the dotted line is not plotted for every instance but for a smaller uniformly sampled subset of instances.}, we simulate the performance of searching around MMSE solution using DI~\cite{di-mimo} and dDI formulations~($16$~users $\times$~$16$~antennas, 16-QAM at 10~dB, 15~dB, 20~dB SNR and 64-QAM at  15~dB, 20~dB, 25~dB SNR) and executing $N_a = 64$ anneals. As demonstrated by the CDFs on the right, the dDI formulation has a much higher probability of finding solutions with no bit errors. In the coupled-plots on the left, the instances are arranged on the y-axis, and the corresponding bit error with DI formulation is on the x-axis. For a subset of instances (not all instances in order to improve the readability of the results), horizontal dotted lines are used to indicate the bit error with dDI formulation. We see that, for a few instances, DI formulation might perform better than dDI, but for a much larger number of instances dDI formulation has significantly better performance. We also note that dDI formulation has a much higher probability of finding a solution with fewer bit errors, and a significantly higher probability of finding the correct solution. Note that, when there is a relatively large amount of noise in the system (10~dB for 16-QAM and 15~dB for 64-QAM), the intrinsic error in the system is itself very high, and while dDI formulation is still better than DI, the gap them is lower. 
\begin{figure}[ht]
  \centering
  \includegraphics[width=1\linewidth]{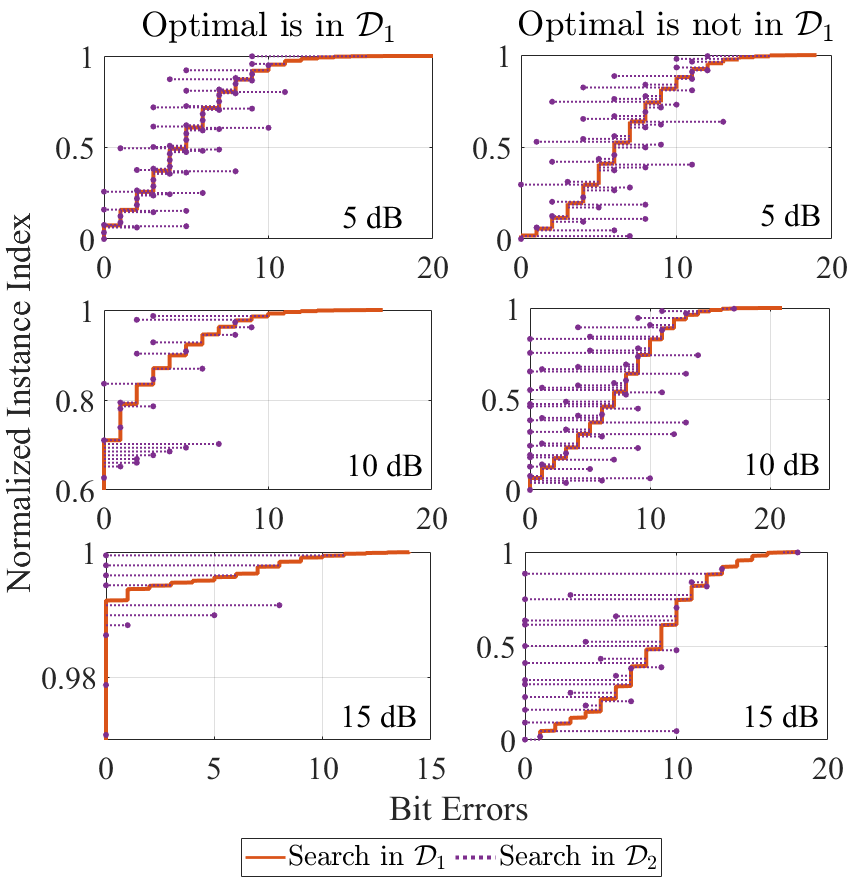}
  \caption{Coupled-plot demonstrating bit errors obtained by searching in different search spaces using the dDI formulations for ($8$ users $\times$ $8$ antennas)~MIMO, 16-QAM system, and $N_a = 32$: while $\mathcal{D}_2$ covers the entire constellation, searching in $\mathcal{D}_1$ can lead to a better performance when the optimal is contained within $\mathcal{D}_1$. However, for scenarios when optimal is not in $\mathcal{D}_1$, a search in $\mathcal{D}_2$ is required.} 
  \label{fig:ss12}
\end{figure}

\begin{figure*}[h]
  \centering
  \includegraphics[width=1\textwidth]{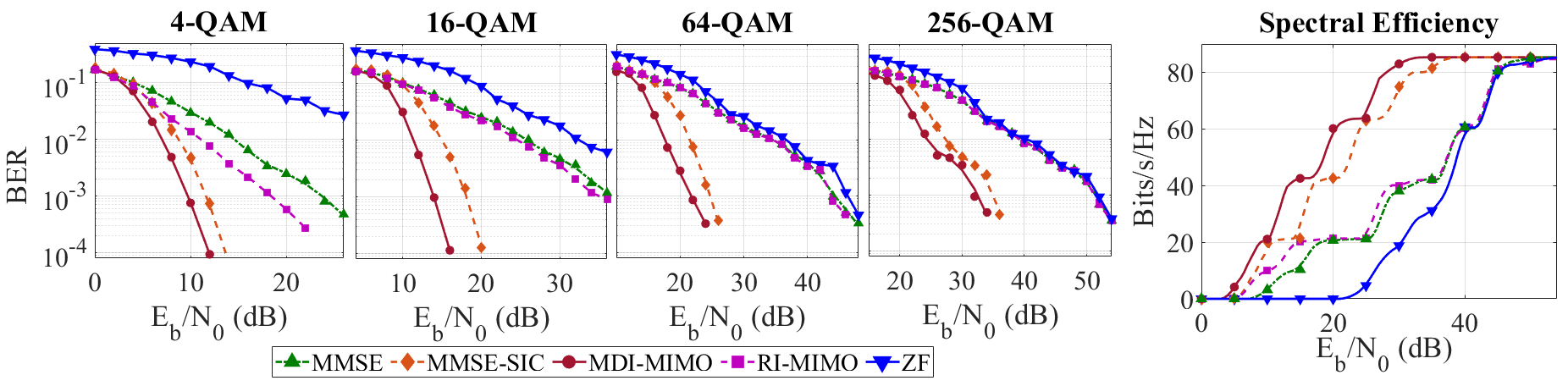}
  \caption{BER and spectral efficiency performance of ($16$ users $\times$ $16$ antennas) Large MIMO on Sionna~\cite{sionna} simulator-based realistic channel model: demonstrating that MDI-MIMO significantly outperforms MMSE, MMSE-SIC, ZF, and RI-MIMO and provides much higher spectral efficiency.} 
  \label{fig:ber16sionna}
\end{figure*}

\begin{figure*}[h!]
  \centering
  \includegraphics[width=1\textwidth]{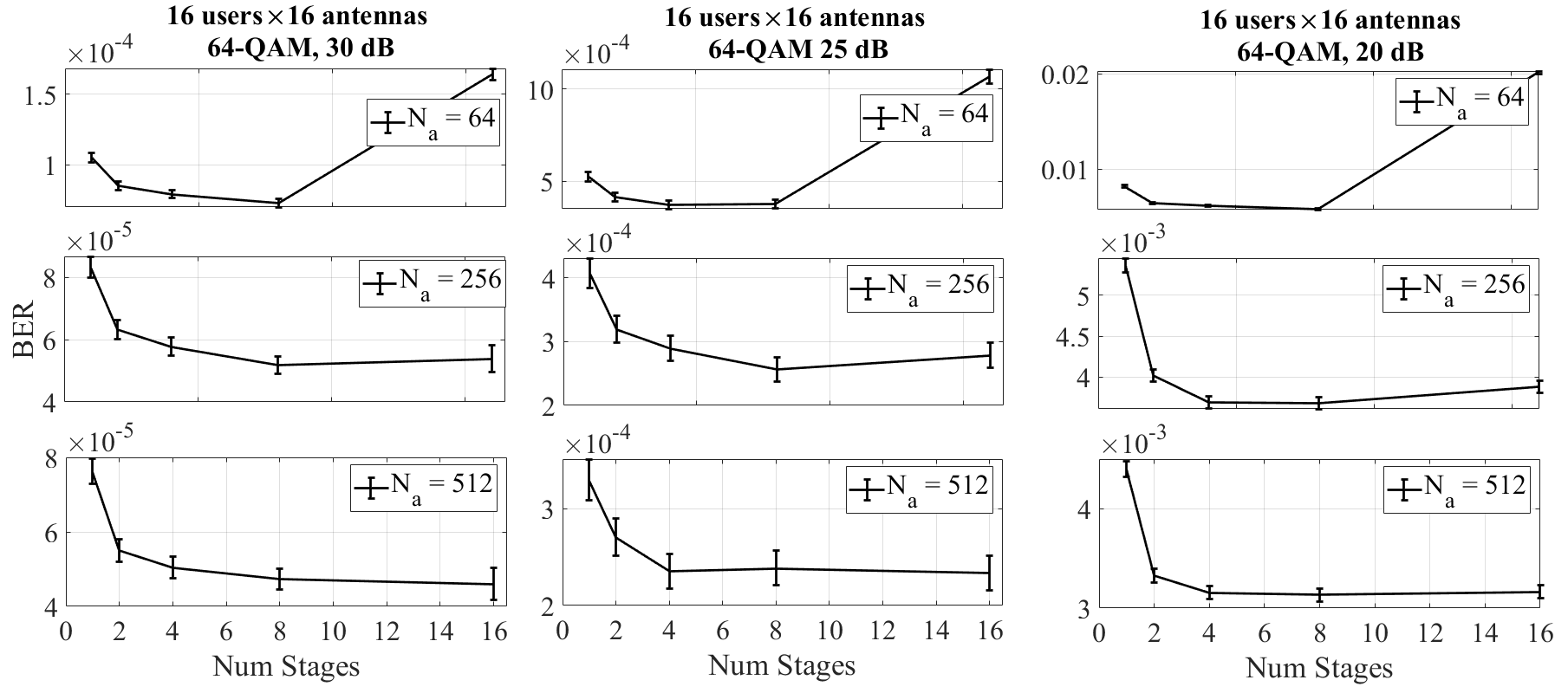}
  \caption{Variation in the BER performance of MDI-MIMO algorithm with a different number of stages while keeping the total number of anneals fixed.}
  \label{fig:ms12}
\end{figure*}

\subsection{Advantages of searching in multiple search spaces}
\label{sec:multipleSearch}
In this section, we will empirically illustrate the necessity for concurrent search in multiple overlapping search spaces. We simulate an ($8$ users $\times$ $8$ antennas, 16-QAM) MIMO system and search around the MMSE-SIC solution using the dDI formulation for $\mathcal{D}_1$ and $\mathcal{D}_2$ (note that $\mathcal{D}_1 \subset \mathcal{D}_2$). We evaluate the `coupled-plot' between the bit errors among the solutions obtained by searching in $\mathcal{D}_1$ and $\mathcal{D}_2$. In these coupled plots, the instances are arranged on the y-axis and the corresponding bit error with search in $\mathcal{D}_1$ is on the x-axis. For several instances, horizontal dotted lines are used to indicate the bit error with search in $\mathcal{D}_2$. 
The underlying trade-off is the following: searching in $\mathcal{D}_1$ requires a lesser number of spin variables than $\mathcal{D}_2$, therefore the problem is less complex, and it is easier for the Ising machine to find the ground state compared to searching in $\mathcal{D}_2$~(which can deteriorate the performance due to lower success probability). However, in the scenarios where the ground state is not in $\mathcal{D}_1$, searching in $\mathcal{D}_2$ or larger is required. 
We see in Fig.~\ref{fig:ss12}, for instances where the optimal solution is in $\mathcal{D}_1$, search in $\mathcal{D}_1$ finds it with very high probability, and for several instances search in $\mathcal{D}_2$ performs worse. However, for the instances when optimal is not in $\mathcal{D}_1$, CIM can find the optimal solution in $\mathcal{D}_2$. Therefore, to cover both these scenarios and achieve the best performance, we should concurrently search over multiple search spaces, like $\mathcal{D}_1$ and $\mathcal{D}_2$ in Fig.~\ref{fig:ss12}~(even though $\mathcal{D}_1 \subset \mathcal{D}_2$). Note that the instances not shown in the coupled-plot (Left) have zero bit errors in both scenarios.

\subsection{Impact of the number of stages in MDI-MIMO}
In this section, we benchmark the performance of MDI-MIMO with respect to the number of stages~($S$). We simulate different scenarios in Fig.~\ref{fig:ms12} by fixing the total number of Anneals~$N_a$. Given a fixed anneal budget~$N_a$, if the number of stages is very high, the total number of anneals per stage is very low, and the performance of individual stages will be bad; on the other extreme, if the number of stages is low, then the advantages of adaptively improving the initial guess and the search radius might be reduced. We observe these trends in Fig.~\ref{fig:ms12}, the performance initially improves with an increase in the number of stages and then starts deteriorating~(as the number of anneals per stage reduces). Empirically we observe that $S = 8$ seems to provide near-best performance.
\begin{figure}
    \centering
    \includegraphics[width=\linewidth]{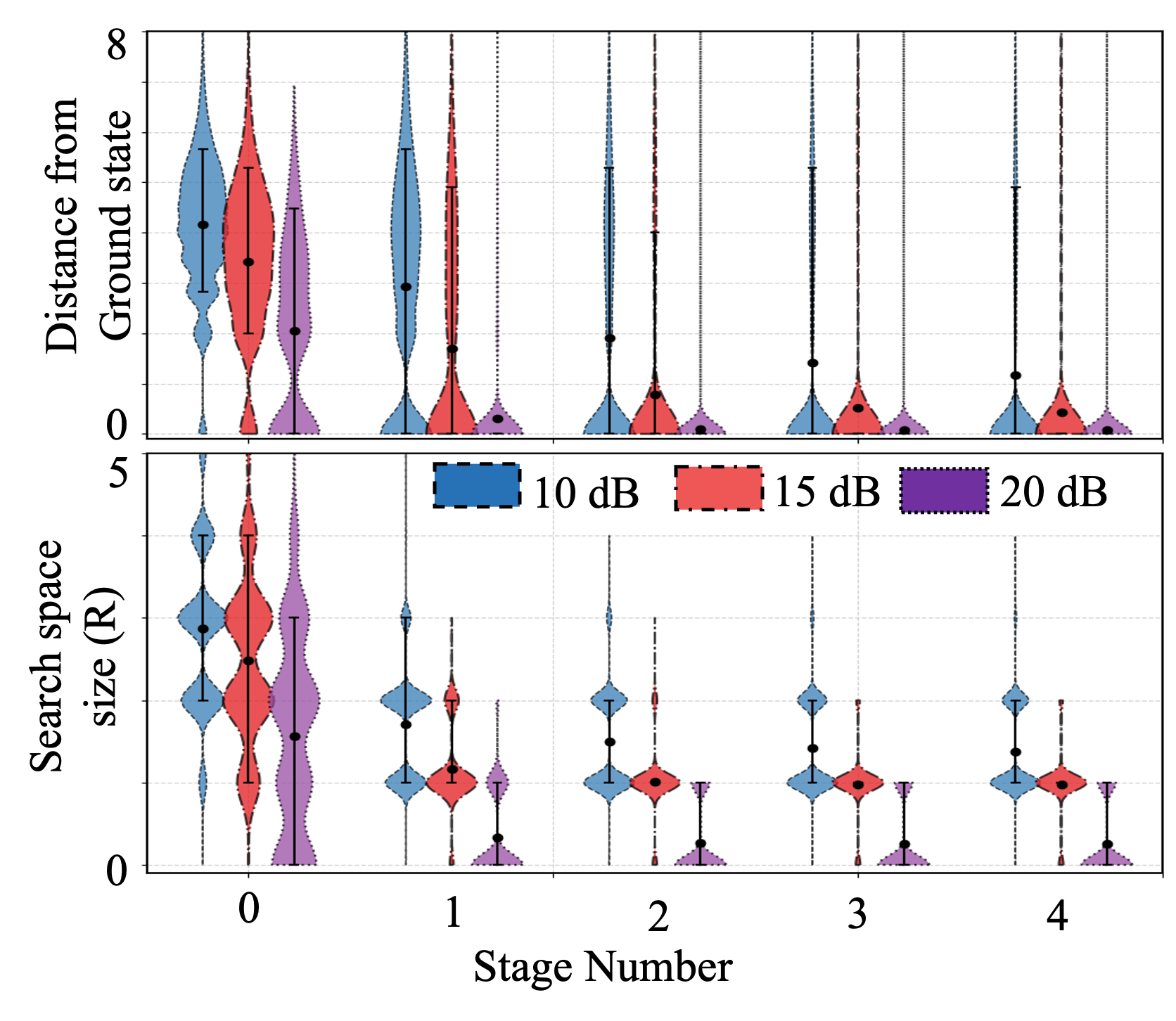}
    \caption{Violin plots for (8 users $\times$ 8 antennas, 16-QAM) MIMO system demonstrating the (top) variation in the distance between the intermediate solution of MDI-MIMO and the ground state with subsequent stages, and (bottom) variation in the required search space determined by MDI-MIMO with subsequent stages.}
    \label{fig:dist_ss_opt}
\end{figure}

The algorithm is designed to re-center the search space at the end of each stage (by modifying the guess used in Eq.~\ref{eq:d_define}) and determine the new appropriate search radius. Since each stage of MDI-MIMO searches for a better solution around the current estimate and improves it, the algorithm tends to zone in on the ground state and hence, we expect the distance between the ground state and the current estimate to reduce in subsequent stages. Further, we expect the required search space to become increasingly smaller. In Fig.~\ref{fig:dist_ss_opt}, we simulate an ($8$ users $\times$ $8$ antennas, 16-QAM, $N_s = 16$) system and we use simplified MDI-MIMO with MMSE as the initial guess (sMDI-MIMO (MMSE)). We plot the mean distance between the output of different stages of MDI-MIMO and the ground state. We also plot the required search space determined at the end of each stage. Note that stage 0  corresponds to the MMSE initialization. We see that the mean distance between the ground state and the intermediate output of MDI-MIMO after each stage steadily reduces with subsequent stages. We also note that the required search space determined by the algorithm also becomes increasingly smaller as MDI-MIMO gets closer to the ground state and with subsequent stages.
\subsection{Performance evaluation on more realistic channel model}
In this section, we evaluate the performance of our methods over a realistic channel model. We generate realistic OFDMA channel traces using the Sionna simulator~\cite{sionna}. We use the 3GPP Urban Macrocell (UMa) channel model and single-sector topology. The base station operates at 2.5 GHz with 128 subcarriers and 15 KHz subcarrier spacing. The base station is equipped with a linear array of 16 horizontally polarized antennas. There are 16 users in the cell, each equipped with one horizontally polarized antenna. All antennas are omnidirectional and users move at a speed of 3~m/s.
This is equivalent to a ($16$ users $\times$ $16$ antennas) MIMO system. We report the average BER over all users and all subcarriers. We see in Fig.~\ref{fig:ber16sionna} that the previously observed performance gains with Rayleigh fading channels also extend to realistic MIMO scenarios. We observe that MDI-MIMO can significantly improve the performance of MMSE and MMSE-SIC, providing up to $2\times$ improvement in throughput over MMSE-SIC and up to 3$\times$ over MMSE. Note that, RI-MIMO~\cite{ri-mimo} only provides performance gains for 4-QAM modulation and its performance becomes similar to MMSE for 16-QAM and higher modulations.
\begin{figure}[h]
    \centering
    \includegraphics[width=\linewidth]{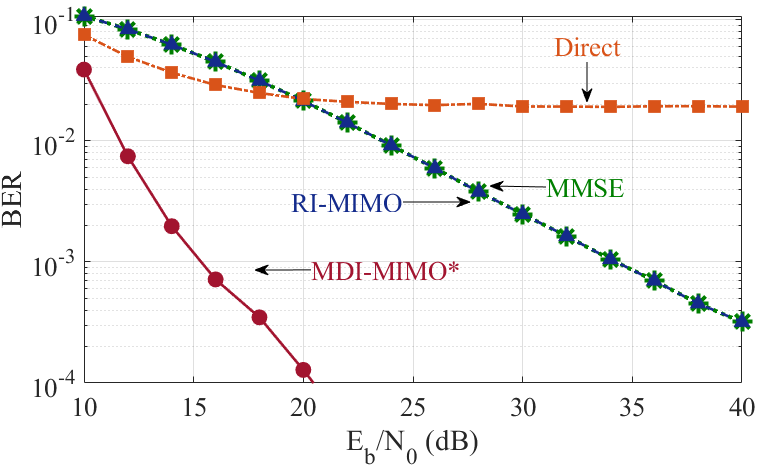}
    \caption{\textbf{Performance with ParaMax:} ($16$ users $\times$ $16$ antennas) MIMO system with 16-QAM. We see that the performance of ParaMax significantly improves even with the simplified variant of MDI-MIMO. } 
    \label{fig:paramax16}
\end{figure}
\begin{figure*}[h!]
  \centering
  \includegraphics[width=1\textwidth]{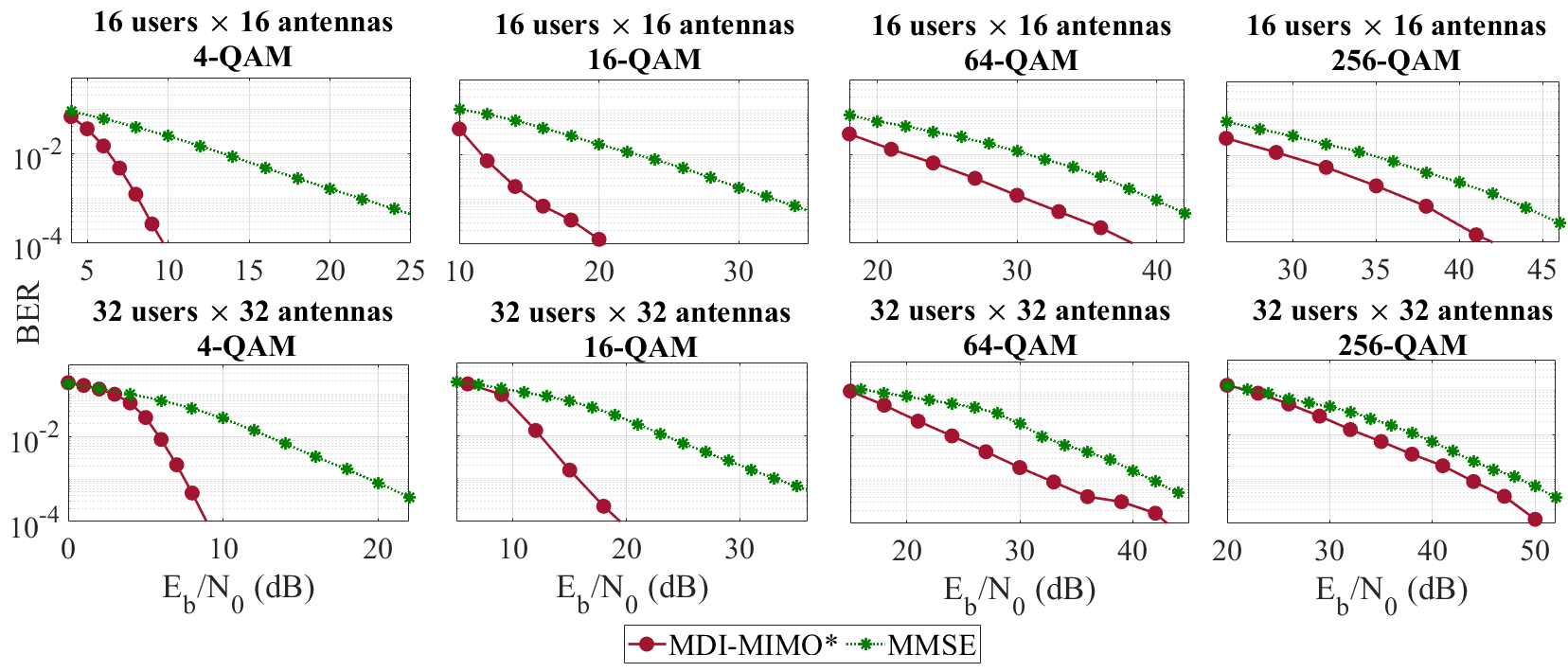}
  \caption{BER performance for ($16$ users $\times$ $16$ antennas) and ($32$ users $\times$ $32$ antennas) Large MIMO: we see that the simplified variant of MDI-MIMO (with ParaMax) significantly outperforms the MMSE detector.} 
  \label{fig:berParamax}
\end{figure*}

\section{Generalization to other Ising machines}
\label{sec:parallelTempering}
In this paper, we have evaluated the performance of our methods using an AHC-based CIM model~\cite{ampCorrectionCIM}. In this section, we will demonstrate the advantages of our methods generalize to other Ising models as well, and they can be used to improve the existing state-of-the-art. 

\subsection{Parallel Tempering and ParaMax decoder}
ParaMax~\cite{minsungParallelTemp} is a \emph{parallel tempering}-based MIMO decoder system that is implemented on the classical platform (CPUs and GPUs). The parallel tempering technique, also known as a replica exchange scheme, is a generic physics-inspired algorithm that can enhance the overall performance of annealing-based optimization methods by helping a system to escape from the local minima using multiple replicas at different temperatures. In ParaMax, a bang-bang parallel tempering approach with only two replicas is applied, where a replica exchange decision is made per Metropolis update~\cite{metropolis1953equation}. The system was able to report near-optimal decoding performance for extreme-large MIMO sizes like ($128$ users $\times$ $128$ antennas)~MIMO with BPSK and 4-QAM modulations.

\subsection{BER performance}
In this section, we evaluate the performance of  MDI-MIMO with Parallel Tempering. We use an extremely simplified variant of MDI-MIMO which uses only MMSE as the starting guess and has a single stage with a fixed search space ($=\mathcal{D}_1$). We will show that the advantages of our methods extend to parallel tempering and significantly improve the state-of-the-art for parallel tempering-based MIMO detection. 

While ParaMax~\cite{minsungParallelTemp} can achieve near-optimal performance for BPSK and 4-QAM, it is unable to provide good performance for 16-QAM and higher modulations: the direct transformation of the ML-MIMO problem into an Ising form fails to reduce the BER beyond a threshold and runs into an error-floor in the BER-vs-SNR characteristics. The RI-MIMO~\cite{ri-mimo} formulation, remedies the error floor but fails to improve the performance for 16-QAM and higher modulation. We see this in Fig.~\ref{fig:paramax16}, for ($16$ users $\times$ $16$ antennas)~MIMO 16-QAM, the direct application of ParaMax leads to the error-floor problem, ParaMax with RI-MIMO~\cite{ri-mimo} remedies the error floor, but provides the same performance as MMSE. However, we see that the simplified variant of MDI-MIMO ($S = 1$ (single stage), fixed search space $R = 1$, and only using MMSE as the initial guess), can significantly improve the performance of existing state-of-the-art; it does not suffer from the error floor problem and provides more than 20dB SNR-gain over the existing state-of-the-art Ising machine based methods. 

We extend our experiments to 4-, 16-, 64-, and 256-QAM modulations and ($16$ users $\times$ $16$ antennas), ($32$ users $\times$ $32$ antennas)~MIMO in Fig.~\ref{fig:berParamax}. We note that, while original ParaMax fails to provide performance improvements for 16-QAM and higher modulations~\cite{minsungParallelTemp}, our methods (with parallel tempering) can provide significant performance improvements over MMSE.

%% file: conclusion.tex
\section{Conclusion and Future Work}
\label{sec:conclusion}
In this paper, we propose our Multi-stage Delta Ising MIMO algorithm~(MDI-MIMO) which performs an adaptive multi-stage search for the maximum likelihood solution of the MIMO problem using a Coherent Ising machine. MDI-MIMO can significantly improve the performance of MMSE and MMSE-SIC, and is the only known Ising machine-based algorithm in the literature~(to the best of our knowledge) that can provide high-performance gains for 16-QAM or higher modulations in Large MIMO systems. Unlike its predecessor DI-MIMO which involves searching in a fixed search radius around an initial guess, MDI-MIMO can adaptively modify the initial guess and predict the appropriate search radius. This allows MDI-MIMO to significantly improve the performance gains, and deal with scenarios where the error in the initial guess is concentrated among a few users. 

Through this work, we have been able to overcome the drawbacks of our predecessors RI-MIMO~\cite{ri-mimo} (works only for 4-QAM or lower) and DI-MIMO~\cite{di-mimo}~(limited performance gains due to fixed search radius), and have demonstrated our performance gains via simulations. As the next step, we plan to focus our efforts on building an FPGA/ASIC-based prototype of the CIM-based MIMO detector, and demonstrating that it can provide good quality solutions while meeting the timing/processing constraints of an LTE system. LTE has very aggressive processing/latency constraints; a typical LTE base station with 20~MHz bandwidth would generate around 8000 MIMO problems every millisecond and need to be solved in 1-2~ms~\cite{ri-mimo}. Meeting such an aggressive timing requirement is a major challenge, and the use of vectorization for numerical integration, multi-threading, and adaptive load-balancing will be key to developing an LTE-compliant prototype.